\documentclass[oldversion]{aa}

\usepackage{graphicx}
\usepackage{txfonts}
\usepackage{amssymb}
\usepackage{longtable}
\usepackage{natbib}
\usepackage{rotating}
\usepackage{morefloats}

\bibpunct{(}{)}{;}{a}{}{,} 

\begin{document}

\titlerunning{{\it Herschel} observations of Hickson compact groups}
\authorrunning{Bitsakis et al.}

\title{{\it Herschel} observations of Hickson compact groups of galaxies:\\ Unveiling the properties of cold dust}

	\author{T. Bitsakis
		\inst{1,2,3}
         \and
     	V. Charmandaris
		\inst{1,4,5}		
	  \and
	P. N. Appleton
		\inst{2}	
       \and
     	T. D\'iaz-Santos
		\inst{6}	
	\and
	E. Le Floc'h
		\inst{7}
	\and 
	E. da Cunha 
		\inst{8}
	\and
	K. Alatalo
		\inst{2}
	\and
	M. Cluver	
		\inst{9}
}

\institute{Department of Physics, University of Crete, GR-71003, Heraklion, Greece
         \and
         NASA {\it Herschel} Science Center, California Institute of Technology, 770 S. Wilson Ave., Pasadena, CA 91125, USA
	\and
	IESL/Foundation for Research \& Technology-Hellas, GR-71110, Heraklion, Greece
	\and
	Institute for Astronomy, Astrophysics, Space Applications \& Remote Sensing, National Observatory of Athens, GR-15236, Penteli, Greece
	\and
	Chercheur Associ\'e, Observatoire de Paris, F-75014,  Paris, France
	\and
	{\it Spitzer} Science Center, California Institute of Technology, 1200 E. California Blvd., Pasadena, CA 91125, USA
	\and
	CEA-Saclay, Orme des Merisiers, Bat. 709, 91191 Gif-sur-Yvette, France
	\and
	Max-Planck-Institut f\"ur Astronomie, K\"onigstuhl 17, 69117 Heidelberg, Germany
	\and
	Department of Astronomy, University of Cape Town, Private Bag X3, Rondebosch, 7701, South Africa
}

\offprints{T. Bitsakis,  e-mail: bitsakis@physics.uoc.gr}

\abstract{We present a {\it Herschel} far-IR and sub-mm study of a sample of 120 galaxies in 28 Hickson compact groups. Fitting their UV to sub-mm spectral energy distributions with the model of da Cunha et al. (2008), we accurately estimate the dust masses, luminosities, and temperatures of the individual galaxies. We find that nearly half of the late-type galaxies in dynamically ``old'' groups, those with more than 25\% of early-type members and redder UV-optical colours, also have significantly lower dust-to-stellar mass ratios compared to those of actively star-forming galaxies of the same mass found both in HCGs and in the field. Examining their dust-to-gas mass ratios, we conclude that dust was stripped out of these systems as a result of the gravitational and hydrodynamic interactions, experienced owing to previous encounters with other group members. About 40\% of the early-type galaxies (mostly lenticulars), in dynamically ``old'' groups, display dust properties similar to those of the UV-optical red late-type galaxies. Given their stellar masses, star formation rates, and UV-optical colours, we suggest that red late-type and dusty lenticular galaxies represent transition populations between blue star-forming disk galaxies and quiescent early-type ellipticals. On the other hand, both the complete absence of any correlation between the dust and stellar masses of the dusty ellipticals and their enhanced star formation activity, suggest the increase in their gas and dust content due to accretion and merging. Our deep {\it Herschel} observations also allow us to detect the presence of diffuse cold intragroup dust in 4 HCGs. We also find that the fraction of 250$\mu$m emission that is located outside of the main bodies of both the red late-type galaxies and the dusty lenticulars, is 15-20\% of their integrated emission at this band. All these findings are consistent with an evolutionary scenario in which gas dissipation, shocks, and turbulence, in addition to tidal interactions, shape the evolution of galaxies in compact groups. }


\maketitle

\section{Introduction}
Since most galaxies are found within large structures, such as groups and clusters, studying how the environment can affect their properties (i.e. stellar populations and morphology) is crucial to understanding their evolution. It is now known that dynamical interactions and the merging of galaxies can affect both their morphologies (induced bars, bridges, tails, and other tidal distortions) and the star formation and nuclear activity \citep[i.e.][]{Struck99}. Nevertheless, galaxy environments can differ depending on the number and the kinematic behaviour of their members. Compact groups of galaxies are systems of several galaxies that display high galaxy densities, similar to those found in the central regions of rich clusters, unlike clusters, they have much lower average velocity dispersions (\citealt{Hickson97} quotes a $\sigma \sim$250 km/s). There are exceptions, such as Stephan's Quintet, where an intruder is colliding with the group with a velocity difference of over 800 km/s \citep{Hickson92}. As a result, galaxies in compact groups can experience a series of strong tidal encounters with other group members during their lifetimes. For these reasons, compact groups are considered ideal systems for studying the effects of dense environments in galaxy evolution. To examine the properties of galaxies in compact groups we used the catalogue defined by \citet{Hickson82}; it consists of 100 groups, containing 451 galaxies, in compact configurations (less than 5 arcmin), within relatively isolated regions where no excess of other surrounding galaxies can be seen. Even though all Hickson compact groups (HCGs) were selected as having four or more galaxies, the original sample was later reduced to 92 groups, since spectroscopic observations revealed the inclusion of interlopers among their members \citep[see][]{Hickson92}.  

\begin{figure*}
\begin{center}
\includegraphics[scale=0.50]{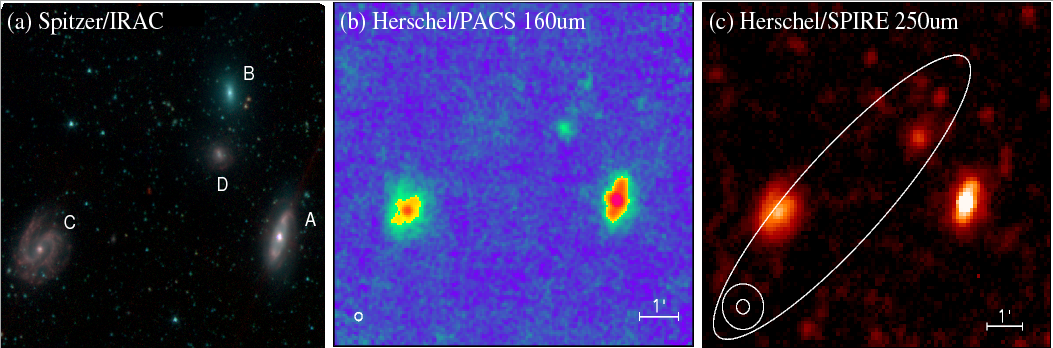}
\caption{Spitzer/IRAC ``true'' colour image of HCG 07 (panel a). The blue channel traces the 3.6$\mu$m emission, the green the 4.5$\mu$m, and the red the 8.0$\mu$m. The {\it Herschel}/PACS 160$\mu$m (panel b), and the {\it Herschel}/SPIRE 250$\mu$m (panel c) images of the same group. In the bottom left corners of panels b and c we include the beam sizes of the {\it Herschel} instruments (11$''$ and 22$''$, respectively). In panel c) we also display for comparison the beams of the {\it AKARI} 160$\mu$m band (70$''$) and {\it IRAS} 100$\mu$m (1.2$'\times$5$'$). }
\label{fig:fig_example}
\end{center}
\end{figure*}

Detailed studies in the past three decades have revealed that HCGs occupy a unique position in the framework of galaxy evolution. They display an excess of elliptical galaxies (relative to the field), with their spiral galaxy fraction nearly half of what it is observed in the field (43\%; \citealt{Hickson82}). \citet{Mendes94} showed that almost half of their galaxies display morphological features of interactions, and \citet{Zepf93} revealed indications of mergers in the irregular isophotes of the elliptical galaxies. In addition, \citet{Ponman96} found hot X-ray gas in $\sim$75\% of the groups, implying that their members reside in a common dark matter halo, while \citet{Desjardins13} show that the X-ray luminosity increases as group HI-to-dynamical mass ratio decreases. However, the hot gas is not in hydrostatic equilibrium and, thus, these systems are not the low-mass analogues of rich groups or clusters. Single-dish radio measurements reveal that galaxies in compact groups are generally deficient in HI with a median mass two times less than what is observed in loose groups \citep{Verdes01}. Furthermore, \citet{MartinezBadenes12} show an increase in the conversion of HI-to-H$_{2}$ caused by the on-going tidal interactions in the spiral galaxies of groups found in early stages of evolution, followed by HI stripping and the decrease in molecular gas because of the lack of replenishment.  Nearly 40\% of HCG members for which nuclear spectroscopy has been obtained display evidence of active galactic nuclei \citep{Shimada00, Martinez10}. 

Although tidal stripping is clearly a strong contender for explaining HI deficiencies, there are cases where the gas seems to have been almost completely stripped from the galaxies, most likely involving both hydrodynamic, as well as pure gravitational, processes. The best studied example is Stephan's Quintet (SQ = HCG92), where HI and molecular studies show that almost all of the cold ISM now resides in the intragroup medium \citep{Williams02,Gao00,Guillard12}. Strong dissipative processes, in addition to gravity, are clearly at work on a large scale in this system, leading to extensive shock heating in the interfaces between galaxies and the intragroup medium, as is evident by the presence of hot X-ray gas, warm molecular hydrogen, and diffuse ionized carbon \citep{Trinchieri05,O'Sullivan09,Appleton06,Cluver10,Appleton13}. 

While systems like SQ seem rare, they are not unique and may represent a rapid hydrodynamic phase in which head-on collisions between the group members can cause rapid sweeping of the gas (e. g. \citealt{Appleton96} for the ring galaxy VII Zw 466) in addition to disruptive tides, conventional mergers, and gas depletion. For example, a recent study of HCG57 by Alatalo et al. (in preparation) also shows a highly disruptive collision between two galaxies in which the molecular and atomic gas is dramatically changed by direct collision. This may be similar to the Taffy system \citep{Condon94}, where a head-on collision has pulled gas into a highly turbulent bridge between the galaxies, heating the molecular gas \citep{Peterson12}. 

\citet{Cluver13} studied the mid-IR spectra of 74 galaxies in 23 HCGs and discovered that more than 10\% of them contain enhanced warm molecular hydrogen emission that, like SQ, may be indicative of shocks and turbulent heating of their ISM. Furthermore, these H$_{2}$ enhanced galaxies (called molecular hydrogen emission galaxies -- MOHEGs by \citealt{Ogle10}) seem to primarily occupy a special place in mid-IR colour space referred to as the ``gap'' by \citet{Johnson07}.  Although not strictly a ``gap'' \citep[see][]{Bitsakis11},  Johnson et al. as well as \citet{Walker12} argue that there is a deficiency of galaxies in the {\it Spitzer} IRAC colour-colour space defined primarily as lying in the range $-0.4<log(f_{5.8}/f_{3.6}) < 0.0$  which they attribute to rapid evolution from the star forming to the quiescent galaxy sequences. In addition, \citet{Walker13} showed that galaxies located in the area of the mid-IR gap, have already transitioned to the optical {\it [g-r]} red sequence. Nevertheless, since these authors do not correct these colours for dust attenuation, some of the galaxies can be found in the red sequence due to extinction. \citet{Cluver13} name the region as the ``infrared green valley'' since it falls between the two main peaks in the mid-IR colour space characterised primarily by star-forming galaxies and galaxies dominated by old stellar populations in compact groups, and as this current paper shows, galaxies in this region tend to be a mix of  dusty red late-type and dusty lenticular galaxies. The discovery that the majority of the galaxies that lie in this region of colour space are also MOHEGs, may lend some credence to the idea that galaxies in the green valley are not just the tail of two separate galaxy populations, but may be galaxies ``in transition'' from an evolutionary perspective. These authors suggest that these shock-excited systems may be heated by collisions with HI clouds liberated into the intragroup medium by previous tidal stripping. SQ could be seen as an extreme example of this process. However, as we see later in this paper, re-accretion of gas by early-type galaxies could also cause evolution of galaxies in the other direction into the green valley. 

In \citet{Bitsakis10,Bitsakis11} -- hereafter $paper~I$ and $paper~II$, respectively -- we obtained the mid-IR fluxes of a sample of 32 HCGs (containing 135 galaxies) to study, for first time, their star formation properties. Using all the multi-wavelength data available, combined with the physically-motivated model of \citet{daCunha08}, we were able to consistently interpret the UV-to-mid-IR spectral energy distribution (SED) of the galaxies in terms of a number of physical parameters, such as their bolometric luminosity, star formation rate, stellar mass, and dust attenuation. We also identified the dynamical state of each group based on the fraction of late-type galaxies they contain, defining as ``old'' (or ``young'') those with more (or less) than 75\% early-type members, and showed that this is consistent with classification based on HI gas content and morphology. We concluded that spiral galaxies in dynamically young groups do not differentiate their star formation properties and UV-optical colours from similar galaxies in other environments, such as in the field and in early stage interacting pairs. On the other hand, the dynamically old late-type galaxies display lower specific star formation rates (sSFR) and redder {\it [NUV-r]} colours than what we see in the field. We suggested that this can be attributed to past galaxy-galaxy interactions and possible minor merging of the group members since the time of its formation, which has increased their stellar masses. Nevertheless, the idea of red late-type galaxies populating the ``green valley'' is not new. \citet{Boselli08} show that ram pressure stripping can remove substantial amounts of atomic gas out of the low-luminosity (L$_{H}<$10$^{9.6}$L$_{H,\odot}$) late-type galaxies entering the Virgo cluster, quenching their star formation and making them display redder colours relatively short timescales ($<$150Myr). Later, this work has been expanded by \citet{Hughes09}, \citet{Cortese09}, and \citet{Gavazzi10} to brighter galaxies in high density environments (i.e. in the clusters of Virgo, Coma, and A1367), showing that the red late-type galaxies display high HI deficiencies and truncation of star formation. Despite the difference in the mechanisms that cause this ``reddening'' (ram pressure in clusters and gravitational stripping in groups) it seems that the effects on stellar populations and star formation activity are similar in compact environments.

In our earlier work the estimates of the far-IR luminosities were based on limited information available at the time for generating SEDs
in the far-IR. Our knowledge of the far-IR was based primarily on 60 and 100$\mu$m measurements from IRAS (which had an $\sim$4 arcmin
beam shape), making it difficult to properly separate galaxies in compact groups that often also lie well within the beam profile. As a result, much of our far-IR knowledge of individual galaxies was necessarily based on extrapolation from the Spitzer 24$\mu$m data points. 

To address this problem, we use in the current paper data taken from several {\it Herschel} Open Time proposals led by our team, as well as other archival data sets from the {\it Herschel} Science Archive\footnote {The Herschel Science Archive is publicly available at http://herschel.esac.esa.int/Science\_Archive.shtml } to make the first detailed analysis of 28 HCG, extending the SED to the far-IR and sub-mm wavelengths. In Section 2, we describe the selection of the samples, used in this work, the data reduction processes, and the UV to far-IR SED modelling of the galaxies. In Section 3, we present the results of our analysis. In Section 4, we discuss  the implications of these findings, and in Section 5 we present our conclusions.

\section{Observations and data analysis}

\subsection{The data}
The sample of Hickson compact groups we present in this paper has been carefully selected from the \citet{Bitsakis11} mid-IR sample,  such that {\it GALEX}/UV, {\it SDSS}/optical, near-IR, and {\it Spitzer}/mid-IR data are available for each individual galaxy. Our sample consists of 28 HCGs (120 galaxies), all observed in the far-IR and sub-mm bands by the {\it Herschel} Space Observatory \citep{Pilbratt10}\footnote{{\it Herschel} is an ESA space observatory with science instruments provided by European-led Principal Investigator consortia and with important participation from NASA.}. As an example of the data in hand, we display the {\it Herschel} 160$\mu$m and 250$\mu$m images of HCG07 in Fig.~\ref{fig:fig_example}, along with a true colour image composite based on {\it Spitzer}/IRAC 3.6, 4.5 and 8.0$\mu$m emission. The comparison of {\it Herschel}, {\it AKARI}, and {\it IRAS} beam-sizes shown in the lower left-hand corner of Fig.~\ref{fig:fig_example}(c) illustrates how significantly {\it Herschel} can improve our knowledge of the spatial distribution and properties of dust in HCG galaxies.

\begin{figure}
\begin{center}
\includegraphics[scale=0.38]{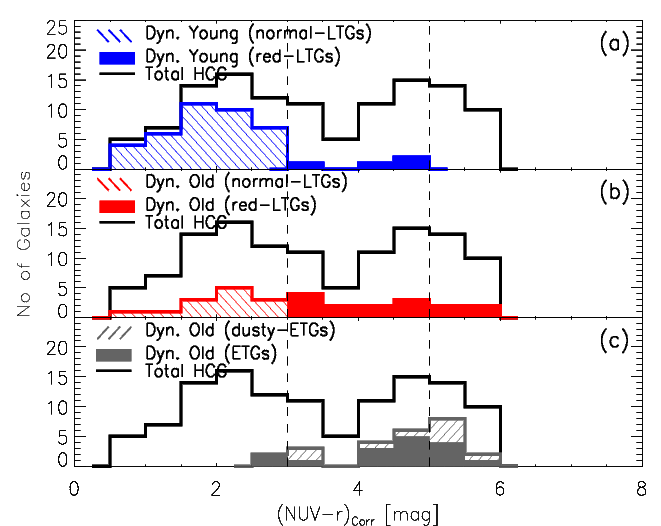}
\caption{Histogram of the extinction corrected {\it [NUV-r]} colours of the dynamically young and old ``blue cloud'' late-type galaxies (normal-LTGs; marked with blue and red shaded lines in panels a and b, respectively) and the dynamically young and old ``green valley'' and ``red sequence''  late-type galaxies (red-LTGs; marked with solid blue and red colours in panels a and b, respectively). The {\it [NUV-r]} colour distributions of the dusty early-type galaxies of the dynamically old groups (dusty-ETGs; marked with a solid grey colour in panel c) and those which were under the detection limit (undetected-ETGs; marked with dashed grey lines in the same panel). The black solid line in all panels represents the total HCG galaxy distribution. The dashed vertical lines separate the different region of the {\it [NUV-r]} colourspace \citep[see][]{Wyder07}, the ``blue cloud'' ({\it [NUV-r]}$<$ 3), the ``green valley'' (3 $<${\it [NUV-r]}$<$ 5), and the ``red sequence'' ({\it [NUV-r]}$>$ 5).}
\label{fig:fig2}
\end{center}
\end{figure}

The observations were obtained between June 2011 and March 2013 (see details in Table~\ref{tab:observations}). From these groups, 18 were observed at $70$ and $160\mu$m by the Photo-detecting Array Camera and Spectrometer (PACS; \citealt{Poglitsch10}), and 6 at the $110$ and $160\mu$m, using scan-map mode at medium scan speed (20$''$/sec). All of these objects were also observed simultaneously at all bands ($250$, $350$, and $500\mu$m) of the Spectral and Photometric Infrared Receiver (SPIRE; \citealt{Griffin10}), using large or small scan-map modes, depending on the source sizes, at the nominal scan speed (30$''$/sec). Additionally, we have recovered the PACS and SPIRE imaging of four more groups (see details in Table~\ref{tab:observations}) from the HSA. The sizes of the instrument beams are $5.5''\times 5.8''$, $6.7'' \times 6.9''$, and $10.7'' \times 12.1''$ for PACS $70$, $100$, and $160\mu$m, and $22''$, $29''$, and $37''$ for SPIRE $250$, $350$, and $500\mu$m bands, respectively. We reduced the Level-1 (timelines) PACS data products using the specialised {\it SCANAMORPHOS} \citep[see][]{Roussel12} and {\it UNIMAP} \citep[see][]{Traficante11} packages, which unlike the standard {\it HIPE v.10} pipeline routines, {\it Naive} and {\it MADmap}, were able to fit and remove the low-frequency $(1/f)$ noise more accurately from our maps. Noise measurements in the final maps indicate typical background uncertainties of 0.15, 0.17, and 0.38 mJy/pixel for the $70$, $100$ and $160\mu$m bands, respectively. The absolute calibration uncertainty is based to the photospheric models of five fiducial stellar standards and has been shown to be 5\% \citep{Balog13}. SPIRE data reduction was carried out using the standard pipeline ({\it DESTRIPPER} in {\it HIPE v.10}). 

Typical background noise in our maps is 0.49, 0.83, and 0.98mJy/pixel for the three SPIRE bands, respectively. Calibration uncertainties are 7\%. The conversion of Jy/beam to Jy/pixel has been done considering beam area values of 465$''^{2}$, 822$''^{2}$ and 1768$''^{2}$ for 250, 350 and 500$\mu$m, respectively. Finally, a colour correction (K$_{4e}$/K$_{4p}$) was applied to all three bands using 0.98279, 0.98344, and 0.97099 correction factors as indicated in the SPIRE Observer's Manual (Sect. 5.2.8). We performed aperture photometry at all bands, using the same apertures we used in the UV, near-, and mid-IR wavelengths. Owing to small differences (3-4\%) in the photometry estimates of the maps reduced by {\it SCANAMORPHOS} and {\it UNIMAP}, we adopted those measured by the former for the purposes of this work. The results are presented in Table~\ref{tab:fluxes}. The total uncertainty in each measurement is given by the sum of the quadrature of the calibration ($\sigma_{cal}$) and photometry ($\sigma_{phot}$) uncertainties. \citet{Ciesla12} show that the first dominates the uncertainties of point-like sources (galaxies with angular diameters less than 20$''$, 29$''$, and 37$''$ at 250, 350, and 500$\mu$m bands of SPIRE) and the second of the more extended sources. The parameter $\sigma_{phot}$ includes the instrumental error (1-2\%), the confusion error due to background sources \citep[see][]{Nguyen10}, and the error of the sky (given by $\sigma_{sky}$=N$_{pixel} \cdot \sigma_{sky,pix}$, where N$_{pixel}$ is the number of pixels in each aperture and $\sigma_{sky,pix}$ the sky error per pixel).

To compare the derived properties and colours of the galaxies in our sample with those of galaxies in the field, we acquired the UV-to-sub-mm fluxes of a similar multi-wavelength dataset for the ``Key Insights on Nearby Galaxies: a Far-IR Survey with {\it Herschel} (KINGFISH; \citealt{Kennicutt11,Dale12}). KINGFISH is a sample of 61 nearby galaxies, observed by {\it Herschel}/PACS and SPIRE and chosen to cover a wide range of galaxy properties and local ISM environments found in the nearby Universe. Most objects are late-type galaxies with angular sizes between 5$'$ and 15$'$. It also contains four early-type galaxies (NGC 855, NGC 1377, NGC 3773, and NGC 4125) with the last one being a LINER. 

\subsection{Galaxy and group classification}
We have used the classifications of \citet{Hickson82} and \citet{deVaucouleurs91}, to identify the morphologies of the galaxies in our sample. From the 120 spectroscopically identified galaxy members in these groups, 59\% are morphologically classified as late-type galaxies (spirals and irregulars, Hubble class T$\ge$1, hereafter LTG) and 41\% are early-type galaxies (ellipticals and lenticulars, with T$\le$0, hereafter ETG). In addition, by applying the classification of the groups' dynamical state, described in $Paper~I$, 13 (46\%) groups are classified as dynamically young (ETG fraction $<$25\%) and 15 (54\%) as dynamically old (ETG fraction $\ge$25\%). 

\begin{figure}
\begin{center}
\includegraphics[scale=0.43]{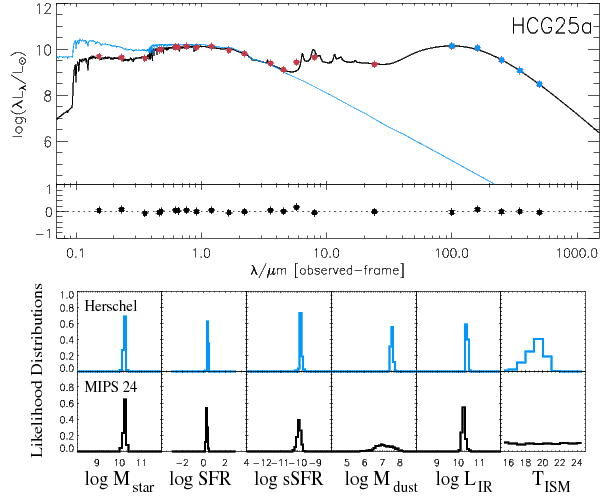}
\caption{{\bf Top panel:} In black the best model fit of the observed SED of HCG25a (a normal-LTG of T=5) shown with red points. The blue line shows the unattenuated spectrum. The blue points correspond to the {\it Herschel}/PACS and SPIRE fluxes. {\bf Bottom panels:} Likelihood distributions of the stellar mass (M$_{star}$), SFR, sSFR, dust mass (M$_{dust}$), dust luminosity (L$_{IR}$), and temperature of the dust in the diffuse ISM (T$_{ISM}$), derived using the UV-to-MIPS 24$\mu$m (in black) and after including the {\it Herschel} bands (in blue).}
\label{fig:fig1}
\end{center}
\end{figure} 

Following the results of $Paper~II$, we further classified our HCG LTGs according to their extinction-corrected UV-optical colours, which are related to their star formation activity. This enabled us to show that the well known bimodality, which should appear between the star forming galaxies of the ``blue cloud'' ({\it [NUV-r]}$<$3) and the quiescent galaxies of the ``red sequence'' ({\it [NUV-r]}$>$5), was not present in the compact group LTGs. More specifically, the distribution of their {\it [NUV-r]} colours not only did not appear to be bimodal, but it also peaked in the intermediate area, the ``green valley'' (3$<${\it [NUV-r]}$<$5). Our results suggested that the LTGs located in the ``green valley'' and ``red sequence'' display older stellar populations, since they built-up their stellar masses possibly as a result of the numerous past interactions. 

In Fig.~\ref{fig:fig2} we have used our model (described in the next section) to reproduce the corresponding Figure 11 from $Paper~II$. We  separated the ``blue cloud'' LTGs (hereafter normal-LTGs) found in dynamically young and old groups from those found in the ``green valley'' and ``red sequence'' areas (hereafter red-LTGs). Finally, we also separated the ETGs depending on whether they were detected in the far-IR and sub-mm bands of {\it Herschel} (hereafter dusty-ETGs) or not (marked simply as ETGs).

\subsection{Estimating the physical parameters of the galaxies}

Using the \citet{daCunha08} MAGPHYS model\footnote{Available at http://www.iap.fr/magphys/} we can consistently compute the IR emission originating in dust in the dense stellar birth clouds and the diffuse ISM of galaxies by proper treatment of the emission from the embedded stellar populations sampled at ultraviolet, optical, and near-IR wavelengths (an example is presented in the top panel of Fig.~\ref{fig:fig1}). This model has been successfully applied to large samples of star forming galaxies from nearby field galaxies, such as the SINGS sample \citep{daCunha08}, to luminous IR galaxies \citep{daCunha10b}. A major advantage of this Bayesian method is that it allows us to compute the probability distribution functions (PDFs) for the various physical parameters and, thus, enables us to easily quantify the physical properties of a given population of galaxies, as well as the uncertainties associated with each parameter (see bottom panels of Fig.~\ref{fig:fig1}). Using MAGPHYS we fitted the SEDs of all galaxies in our sample and in the comparison sample and estimated their star formation and dust properties. The results are presented in Tables~\ref{tab:properties} and \ref{tab:KINGFISH}, respectively. 

\begin{figure}
\begin{center}
\includegraphics[scale=0.38]{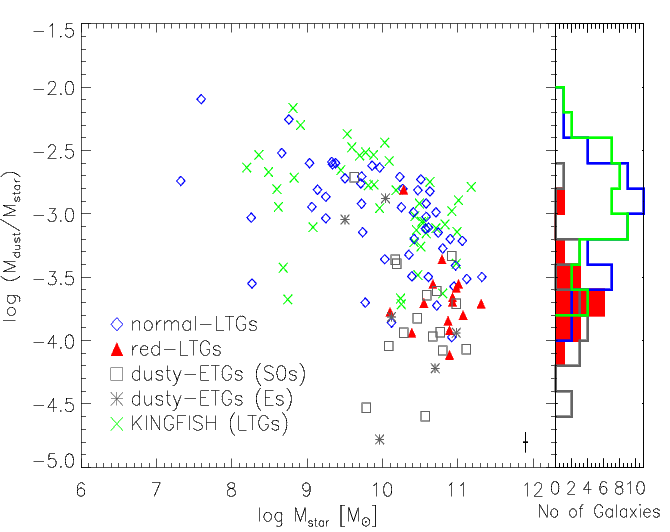}
\caption{Dust-to-stellar mass ratios as a function of the stellar mass for the normal-LTGs in dynamically young and old groups (denoted by empty blue diamonds), the red-LTGs (marked with filled red triangles), the dusty ETGs (S0s are marked with grey squares and Es with grey stars), and the KINGFISH LT galaxies (marked with green x's). Typical uncertainties are indicated using error bars in the bottom right corner.}
\label{fig:fig4}
\end{center}
\end{figure}

As we mentioned in the introduction, up until the launch of {\it Herschel}, the lack of deep, high spatial-resolution far-IR imaging of the groups, essential for resolving each member galaxy, has caused a serious limitation in understanding the dust properties (such as the dust masses, luminosities, and temperatures) of the group members. In the bottom panels of Fig.~\ref{fig:fig1} we present an example of the derived PDFs of the stellar mass, star formation rate (SFR), specific-SFR (sSFR), dust mass and luminosity (M$_{dust}$ and L$_{IR}$, respectively), and cold dust temperature (T$_{ISM}$) of the galaxy HCG25a. We plot the histograms of the parameters derived using UV-to-MIPS 24$\mu$m data and the ones that also include the {\it Herschel}/PACS and SPIRE bands. We note that, as expected, the uncertainty in the PDFs of the stellar mass and SFR of HCG25a is reduced by a factor of 1.12 and 2.7, respectively, without a major change in the most likely value. However, the corresponding values of the luminosity and the mass of the dust change significantly ($\sim$65\%), and their PDFs are better constrained by a factor of $\sim$10. Finally, the average temperature of cold dust, which was completely unconstrained, can now be accurately estimated.

\begin{figure}
\begin{center}
\includegraphics[scale=0.38]{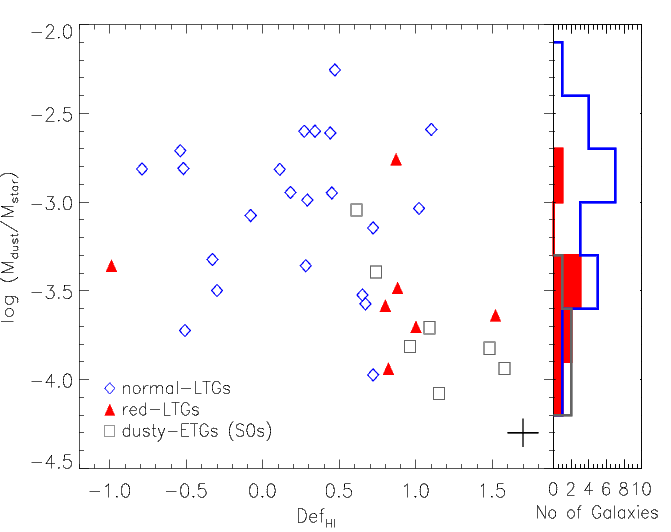}
\caption{Dust-to-stellar mass ratios of the galaxies in our sample versus the HI gas deficiencies. Symbols as in Fig. 4. Typical uncertainties are indicated using error bars in the bottom right corner.}
\label{fig:fig5}
\end{center}
\end{figure}

Since the stellar masses and the star formation rate values of the galaxies are slightly modified after including the far-IR data, we checked whether our findings from $Paper~II$ are consistent with the estimations of this work. Nevertheless, to make that comparison we have to take one more change into account. We note that the current version of MAGPHYS uses the \citet{Bruzual03} stellar-population synthesis models (hereafter BC03), rather than the Charlot \& Bruzual (2007) libraries (hereafter CB07), which were used at the time we estimated those parameters in $Paper~II$. \citet{Zibetti12} show that the latter overestimate the impact of the thermally pulsating asymptotic giant branch (TP-AGB) stars, artificially increasing the near-IR fluxes of the stellar evolution models, which in reality are closer to the BC03 model estimates. To investigate whether these biases affected our previous findings, we re-examine the distributions of the these parameters as derived in our present study. We notice that the stellar masses of the dynamically old red-LTGs have a median value of $\sim$7.76$^{+9.74}_{-4.32}$$\times$10$^{10}$M$_{\odot}$. This is significantly higher than the median masses of the normal-LTGs found in the groups and the field (displaying P$_{KS}$\footnote{The Kolmogorov-Smirnov test probability. If less than 1\%, the two distributions can be considered as significantly different.} less than 0.1\% in all cases), which have median stellar masses of $\sim$1.70$^{+5.17}_{-1.27}$$\times$10$^{10}$M$_{\odot}$  and 1.51$^{+10.0}_{-1.31}$$\times$10$^{10}$M$_{\odot}$, respectively. These results are in excellent agreement with our previous findings. Examining the SFR distributions of the red-LTGs in dynamically old groups, we notice that they display significantly lower SFRs (P$_{KS}\sim$10$^{-4}$\%) than the normal-LTGs in the other subsamples. More specifically, the former are found to have up to an order of magnitude lower SFRs (with a median SFR of $\sim$0.09$\pm$0.21 M$_{\odot}$ yr$^{-1}$) than normal-LTGs of the same mass range found in groups and the field (with 1.05$\pm$0.22 M$_{\odot}$ yr$^{-1}$ and 0.96$\pm$0.22 M$_{\odot}$ yr$^{-1}$, respectively). Additionally, we find that the SFR per unit mass (or specific SFR; sSFR) of the red-LTGs displays significantly lower values (0.37$^{+1.06}_{-0.28}$$\times$10$^{-11}$ yr$^{-1}$) than those of the normal-LTGs in groups and the field (6.02$^{+16.4}_{-5.18}$$\times$10$^{-11}$ yr$^{-1}$ and 5.37$^{+19.5}_{-5.02}$$\times$10$^{-11}$ yr$^{-1}$, respectively), which is precisely what we concluded in $Paper~II$. 

\section{Results}

\subsection{The dust masses of the HCG late-type galaxies}
Interstellar dust is produced during the last stages of stellar evolution. The outflows of stellar winds and supernova explosions enrich the interstellar medium with dust grains of various sizes. There the dust grains are mixed well with the atomic gas, leading to typical gas-to-dust mass ratios of $\sim$150-200 \citep{Draine07,Thomas04}. Once in the ISM, dust grains can grow by accretion, but are also subject to destruction (from interstellar shocks or hot X-ray plasmas). There are several mechanisms that can destroy dust grains, such as the thermal sputtering and grain-grain collisions \citep{Draine03}. \citet{Cortese12} showed that in a ``closed-box'' model, the dust-to-stellar mass ratio decreases with decreasing gas fraction. The reason is that as galaxies build up their stellar mass, consuming their available gas, they decrease their star formation activity, and as a consequence they replenish the ISM with new dust and slow down. This is also consistent with the results of \citet{daCunha10a}, who found that the dust-to-stellar mass ratios anti-correlate with the sSFRs and, as a result, with the {\it [NUV-r]} colours in LTGs. 

\begin{figure}
\begin{center}
\includegraphics[scale=0.39]{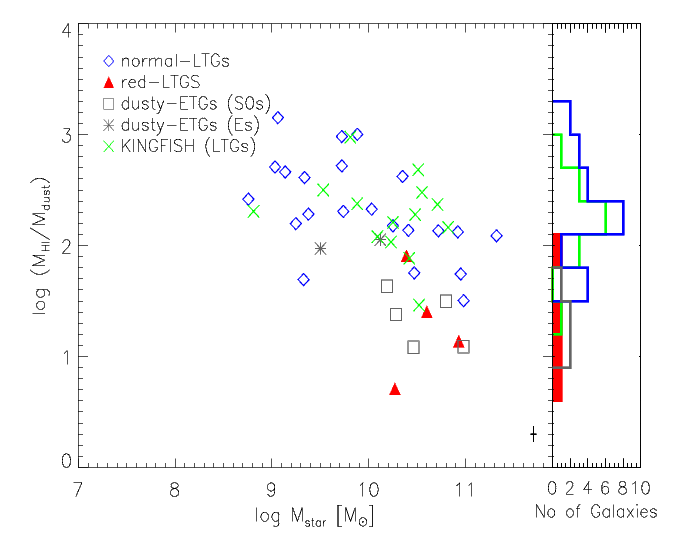}
\caption{The atomic gas-to-dust mass ratios as a function of the stellar mass. Symbols as in Fig.~\ref{fig:fig4}. Typical uncertainties are indicated using error bars in the bottom right corner.}
\label{fig:fig6}
\end{center}
\end{figure}

This anti-correlation between the dust-to-stellar mass ratios and the stellar masses is also observed in the HCG galaxies of our sample (see Fig.~\ref{fig:fig4}). We can see that the most massive normal-LTGs, with M$_{star}>$10$^{10}$M$_{\odot}$, display nearly 30\% lower dust-to-stellar mass ratios compared to the less massive ones.  However, what is worth noticing here is that for a given stellar mass, the red-LTGs display significantly lower (P$_{KS}$$<$10$^{-6}$\%) dust masses  than the normal-LTGs in HCGs and in the field of the same mass range, with typical dust-to-stellar mass ratios of 1.9$\pm$1.0$\times$10$^{-4}$, 9.5$\pm$3.8$\times$10$^{-4}$, and 10.5$\pm$5.4$\times$10$^{-4}$, respectively. In the upper panel of Table~\ref{tab:mass_diffuse_dust} we present the fraction of dust mass the red-LTGs are missing with respect to the average dust-to-stellar mass ratio shown by normal-LTGs. This fraction ranges from $\sim$30 to 80\%, with the only exceptions galaxies HCG40e and HCG95d, whose dust-to-stellar mass ratios are similar to those of normal-LTGs. 

\begin{figure}
\begin{center}
\includegraphics[scale=0.39]{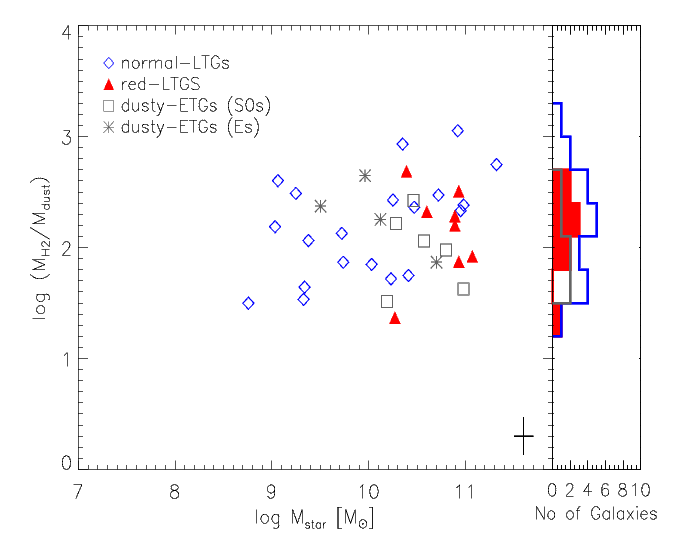}
\caption{Molecular gas-to-dust mass ratios as a function of the stellar mass. Symbols as in Fig.~\ref{fig:fig4}. Typical uncertainties are indicated using error bars in the bottom right corner.}
\label{fig:fig7}
\end{center}
\end{figure}

\citet{Cortese12} show that most HI-deficient galaxies in the Virgo cluster were those with the lowest dust-to-stellar mass ratios. Using the results from \citet{Verdes01}, we plot in Fig.~\ref{fig:fig5} the dust-to-stellar mass ratios of the LTGs in our sample versus their HI gas deficiencies, defined as Def$_{HI}$ = log[M(HI)$_{pred}$] $-$ log[M(HI)$_{obs}$], where the predicted HI masses are estimated from the optical luminosity L$_{B}$ as explained in \citet{Haynes84}. Despite the small number of available galaxies with measurements of HI deficiencies, red-LTGs tend to be more deficient than the normal-LTGs. What mechanisms  are responsible for stripping gas and dust in these galaxies? \citet{Verdes01} and \citet{Borthakur10} suggest that the atomic gas was removed from these galaxies as a result of tidal stripping. Other processes that could strip gas from galaxies have also been considered. \citet{Rasmussen08} show that although many evolved HCGs contain extended soft-X-ray emission, ram-pressure stripping is insufficient and cannot explain the observed HI deficiencies in all but the most X-ray bright groups. Another possibility was explored by \citet{Cluver13}. These authors noted that HI removed from previous tidal interactions would remain confined to the intragroup medium, (as suggested by deep GBT observations by \citealt{Borthakur10}), and these clouds could continue to collide with the galaxies as they move through the group, perhaps further stripping material from the galaxies. HCG92 (StephanÕs Quintet) is one example where a galaxy has collided with previously tidally stripped HI plume, creating a powerful molecular-rich shock. In that system almost all the gas is stripped from the galaxies and lies in the IGM (see \citealt{Guillard12,Appleton13}).

In Figs.~\ref{fig:fig6} and \ref{fig:fig7}, we plot the atomic and molecular gas-to-dust mass ratios versus the stellar masses of the galaxies in our sample, respectively. Using the measurements of \citet{Verdes01}, \citet{MartinezBadenes12} and \citet{Leon98}, we notice that the red-LTGs exhibit the lowest atomic gas-to-dust mass ratios ($\sim$14) among the overall population of LTGs ($\sim$190). On the other hand, when we examine the molecular gas-to-dust mass ratios, we cannot see any difference between the different LTG sub-samples (M$_{H2}$/M$_{dust}$$\sim$200). \citet{Cortese12} found similar results for galaxies in the Virgo cluster and suggest that it is easier for the dynamical interactions to strip the atomic gas out of the galaxies' disks, rather than their molecular gas and interstellar dust. They attributed this to the fact that conversely to the HI disk, which extends nearly two times farther than the stellar one, the dust and the molecular gas are more centrally concentrated \citep[i.e. see][]{Thomas04}. This could be caused by metallicity gradients in the outskirts of galaxies. Galaxy interactions will strip the HI in these regions, which is poorer in interstellar dust. However, dust plays a key role in the formation of H$_{2}$, thus keeping the molecular gas-to-dust mass ratio constant. 

\begin{figure}
\begin{center}
\includegraphics[scale=0.38]{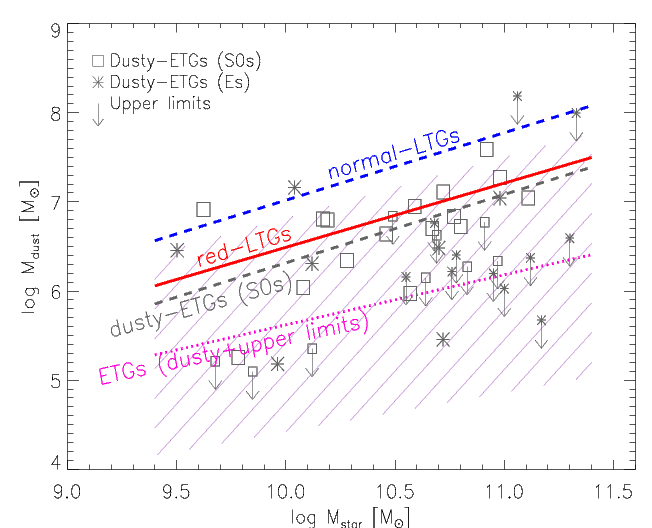}
\caption{Dust versus the stellar masses for the early-type galaxies in groups. Galaxies under the detection limit of PACS and SPIRE bands are marked with upper limits. Lenticulars (S0's) are marked by grey squares, ellipticals (E's) by grey stars. The blue dashed and the red solid lines indicate the linear fits of the ``blue cloud''  and the red-LTGs, respectively. The grey dashed line is the linear fit of the dusty lenticulars; we did not include a fit for the dusty ellipticals due to the complete absence of any correlation between their dust and stellar masses. The magenta dotted line indicates the linear fitting of all ETGs including the ones that they are upper limits, as estimated from the Buckley-James survival analysis method. The magenta shaded area corresponds to the uncertainties of the survival analysis method.}
\label{fig:fig8}
\end{center}
\end{figure}

\subsection{Dust masses of HCG early-type galaxies}
The detection rate of ETGs in the far-IR is noticeably different between dynamically young and old groups. In our sample, none of the seven ETGs in dynamically young groups were detected by {\it Herschel}, and yet 40$\%$ of the 43 ETGs in dynamically old groups were detected in all bands of {\it Herschel} (hereafter dusty-ETGs). Almost 70\% of these ETGs are classified as lenticulars and 30\% as ellipticals. In Figs.~\ref{fig:fig4}, \ref{fig:fig6}, and \ref{fig:fig7} we present the dust-to-stellar, as well as the atomic and molecular gas-to-dust mass ratios of these dusty-ETGs. We can see that these galaxies display very similar ratios to those of red-LTGs, also according to statistical tests having P$_{KS}>$20\% in all cases. 

In Fig.~\ref{fig:fig8}, we plot the dust mass as a function of the stellar mass of the dusty ETGs, but this time by separating them into lenticulars (0$\ge$T$\ge$-3) and ellipticals (T$\le$-4). Galaxies that were not detected in all bands of  $Herschel$ are marked with upper limits, since, as we showed in Sect.~2.3, these bands are crucial for  estimating of their dust masses. The large dispersion in the upper limits is a result of the differences in integration times and distances to these objects. The dust masses of the dusty lenticulars appear to correlate with their stellar masses (R=0.58\footnote{The Pearson correlation coefficient measures the linear correlation between two variables, where 1 is the total linear correlation and 0 is no correlation.}), which is consistent with what we have seen in the various LTG sub-samples (R=0.74 and 0.86 for the red and normal-LTGs, respectively). This could be attributed to the fact that as galaxies evolve and increase their stellar content, more and more stars enrich their interstellar medium with dust. In addition to what we have seen in the previous figures, dusty lenticulars display similar dust-to-stellar mass ratios to those of the red-LTGs of the same stellar mass range (1.5$\pm$0.5$\times$10$^{-4}$ and 1.9$\pm$1.0$\times$10$^{-4}$, respectively), described in Sect.~3.1. The fraction of the dust deficit with respect to their stellar mass is presented in the medium panel of Table.~\ref{tab:mass_diffuse_dust}. Is it possible that these galaxies are somehow connected from an evolutionary perspective? 

\begin{figure}
\begin{center}
\includegraphics[scale=0.38]{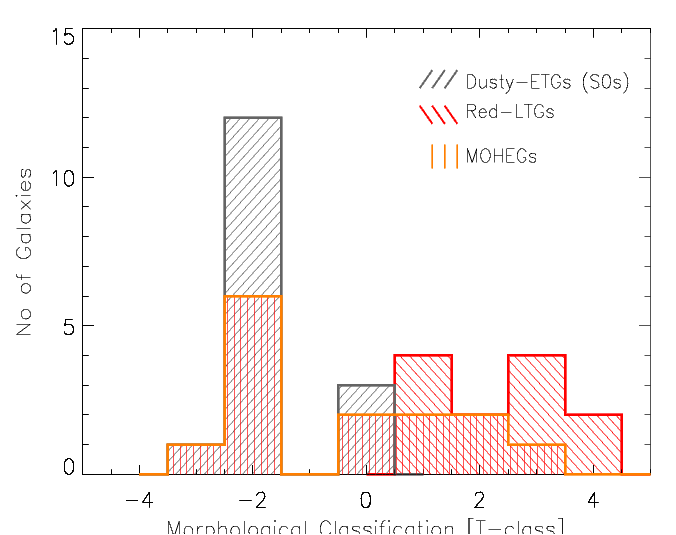}
\caption{Morphological classification (T-class) of the red late-type galaxies (in red), the dusty lenticulars (in grey) and the galaxies classified as MOHEGs (in orange).}
\label{fig:fig9}
\end{center}
\end{figure}

\begin{figure*}
\begin{center}
\includegraphics[scale=0.46]{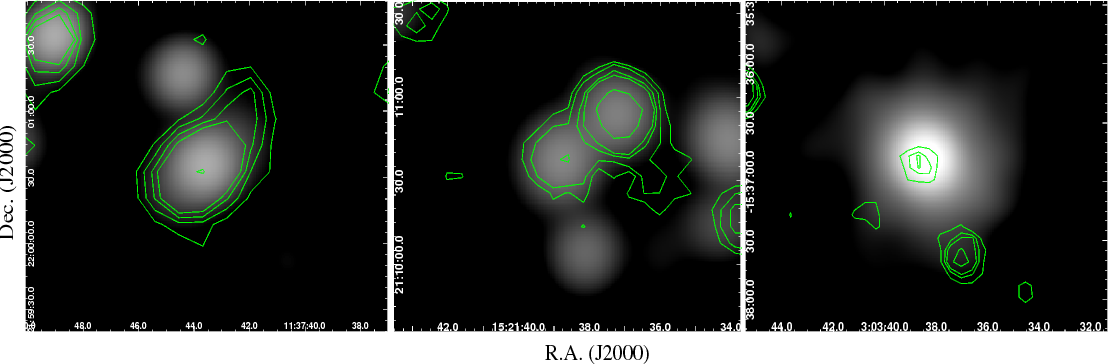}
\caption{{\it Spitzer}/IRAC 3.6$\mu$m images of normal-LTG HCG95c, red-LTG HCG57b (second panel), and dusty-ETG HCG22a (third panel) convolved to the size of the beam of {\it Herschel}/SPIRE 250$\mu$m. The contour maps are the {\it Herschel}/SPIRE 250$\mu$m of the same groups. The five contours correspond to 3, 5, 7, 9, and 20-$\sigma$.}
\label{fig:fig10}
\end{center}
\end{figure*}

Examining their stellar masses and specific star formation rates we see that they are very similar to those of the red-LTGs (3.89$^{+6.34}_{-2.41}$$\times$10$^{10}$M$_{\odot}$ and 0.14$^{+0.40}_{-0.10}$$\times$10$^{-11}$ yr$^{-1}$, respectively), and significantly different from those observed in the normal-LTGs (see the conclusions of $paper~II$). They both primarily occupy the UV-optical and IR green valley, with the dusty lenticulars being somewhat redder. Studying the morphological classifications of the HCG galaxies in Fig.~\ref{fig:fig9}, we can see that the red-LTGs have a T-class between 1-4, while dusty lenticulars have T$\le$0. Approximately 60\% of the red-LTGs and dusty lenticulars -- for which we had available mid-IR spectra -- were identified as MOHEGs by \citet{Cluver13}. These results are consistent with the simulations of \citet{Bekki11}, where the repetitive slow encounters experienced by a spiral galaxy in the compact environment of a group will result in the build-up of its bulge mass, by fuelling gas towards the centre. This will initially increase the stellar mass of the galaxy, and possibly its migration towards redder colours, due to the ageing of the stellar populations. Eventually, after many encounters the increase in the central bulge will transform the galaxy into a lenticular. Therefore, based on these findings we suggest that the red-LTGs and the dusty lenticulars are two contiguous galaxy populations in morphological and colour transformation from the green valley to the quiescent galaxy sequence. It is interesting that these same systems also tend to be the same galaxies as shown by \citet{Cluver13} to contain unusually large warm molecular hydrogen emission. This might suggest a relationship between shocked molecular hydrogen and possible star formation quenching \citep[see][]{Konstantopoulos13}. 

Contrary to what we have seen in the various LTG samples and the dusty lenticulars, the dust masses of the dusty ellipticals (see Fig.~\ref{fig:fig8}) do not correlate with their stellar masses (R=0.09). One could argue that these galaxies might have experienced stronger winds and outflows than dusty lenticulars, which as a result expelled large amounts of dust out of their interstellar medium. Even though this could explain the very low dust-to-stellar mass ratios of a couple of these dusty ellipticals, it does not agree with the fact that many of these galaxies exhibit ratios similar to or even higher than those of the dusty lenticulars and the red-LTGs. In addition, we can see that their sSFRs are significantly higher than those of the remaining ellipticals in groups with no detected far-IR emission (1.35$\times$10$^{-12}$ yr$^{-1}$ and $\le$0.48$\times$10$^{-12}$ yr$^{-1}$, respectively). This implies that some dusty ellipticals may be experiencing a late star formation phase, possibly because they accrete a gas-rich dwarf galaxy, which would make them appear as rejuvenated. In addition, their enhanced dust content could be a result of accretion from other group members and/or the intragroup medium, as well as from the last stage of star formation activity they are going through. Similar results have been observed by \citet{diSerego13} for elliptical galaxies in the Virgo cluster, as well as in galaxy pair Arp 94, where the elliptical galaxy NGC3226 increased its dust content from its interaction with spiral galaxy NGC3227 (Appleton et al. - in prep.).

\subsection{Diffuse cold dust in the intragroup medium}
So far our results suggest that a significant fraction of the interstellar dust of the red-LTGs and dusty-ETGs was stripped out into the intragroup medium,  but where is the missing dust? It is possible that some fraction of it could be still located near these galaxies. To explore this possibility, we have masked their stellar disks and measured the 3-$\sigma$ {\it Herschel}/SPIRE 250$\mu$m extended emission, typically brighter than $\sim$20.1 mag arcsec$^{-2}$. To define the size of the stellar disk to be masked we convolved the {\it Spitzer}/IRAC 3.6$\mu$m image to the size of the beam of SPIRE 250$\mu$m and then carefully calculated the 3-$\sigma$ near-IR isophotal contours around each source down to a typical 3.6$\mu$m brightness of $\sim$24.5 mag arcsec$^{-2}$. The results are presented in columns 2-5 of Table~\ref{tab:mass_diffuse_dust}. In the majority of red-LTGs, 20\% of their total 250$\mu$m flux on average is extended and originates outside their stellar disks. The same holds for only a third of the dusty-ETG when an average of 18\% of their 250$\mu$m emission is extended.  No extended emission is seen in normal-LTG, with HCG95c (see first panel of Fig. 10) and HGC100c being the only exceptions. In the second panel of Fig.~\ref{fig:fig10} we present an example of extended emission in red-LTG HCG57b. On the other hand, there are nine galaxies -- mostly ellipticals -- where, in addition to the very high dust deficiencies they display, the extent of their dust disk is more than 20\% smaller compared to their stellar one (see last panel of Fig.~\ref{fig:fig10}). Examining the SPIRE 250$\mu$m contour maps of galaxies HCG15f and HCG95c (see Appendix A) we notice ``dusty'' bridges connecting them with their companions, HCG15d and HCG95a, which are in both cases the most massive galaxies in their groups. Moreover, the far-IR contours of theHCG16b and HCG79b reveal that their dust distributions are concentrated towards the closest members in their groups, leaving large areas of their stellar disks without dust. 

\begin{figure}
\begin{center}
\includegraphics[scale=0.39]{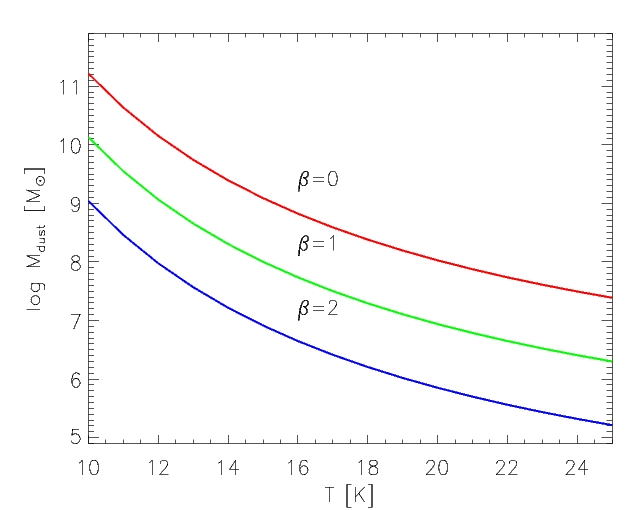}
\caption{An example of the dependence in the estimation of diffuse dust (Eq.~1) by the dust emissivity index ($\beta$) and the dust temperature (T) in HCG92.}
\label{fig:fig11}
\end{center}
\end{figure}

To estimate the amount of mass of the extended dust (M$_{ExtD}$) in these systems, we use of a modified black-body, as presented in \citet{Taylor05}:\\
\begin{equation}
M_{D}[M_{\odot}]=3.24\times10^{-44} {fv[Jy] D_{L}^{2}[cm^{2}] exp({0.048v_{obs}[GHz](1+z)\over T[K]}] \over (1+z)\cdot({v_{obs}[GHz](1+z)\over 250})^{(\beta+3)}}
\end{equation}
where $\beta$ is the dust emissivity index and T the expected dust temperature. The dust emissivity index varies between 0-2 depending on the size of the various dust grains. In Fig.~\ref{fig:fig11}, we notice that the dust mass of HCG92 can change by almost two orders of magnitude depending on the assumptions we make for $\beta$ and T. Following the most recent {\it Herschel} and {\it Planck} observations, we adopt a $\beta$=1.5, which was shown to represent the dust emissivity better in normal star forming galaxies \citep{Boselli12, Planck11}. In addition, a temperature range 15-20K was chosen to estimate the dust mass. As an upper limit we used the ISM cold dust temperature of our galaxies derived from the model and as a lower limit the estimated temperature of the cold dust stream $\sim$15kpc from the centre of NGC5128 \citep{Stickel04}.The results are presented in the last column of Table~\ref{tab:mass_diffuse_dust}. As we see, the amount of extended dust we detect near our galaxies is not enough to explain the high dust deficiencies of these systems (presented in col. 6 of the same table). Therefore, the question remains: where is the remaining dust? Is it located in the intragroup medium and if so, in what form, diffuse or clumpy?

In a preliminary analysis, discussed in $Paper~II$, we attempted to investigate the contribution of each individual galaxy to the total far-IR emission of its group, as measured by IRAS \citep{Allam96}, and identified several groups (mostly classified as dynamically old) in which diffuse cold intragroup dust might be present. Applying the same technique for the galaxies in our present sample, we have estimated the combined ``synthetic'' IRAS 60 and 100$\mu$m flux densities of all group members, as predicted by our SED model, and compared them with the integrated values of each group measured by IRAS. We find that in at least in four groups (HCG16, HCG54, HCG79 and HCG92) there is an excess greater than 3$\sigma$ in the integrated IRAS emissions compared to what is measured by {\it Herschel} (see Table~\ref{tab:diffuse}); the diffuse dust properties of HCG92 are presented in more detail in Guillard et al. (in prep.). Examining the IRAS scans (using SCANPI, an interactive software tool for inspecting the calibrated survey scans from IRAS) we confirmed the complete absence of any peculiar scans in the IRAS maps, which could be attributed to Galactic cirrus emission. 

\begin{figure*}
\begin{center}
\includegraphics[scale=0.75]{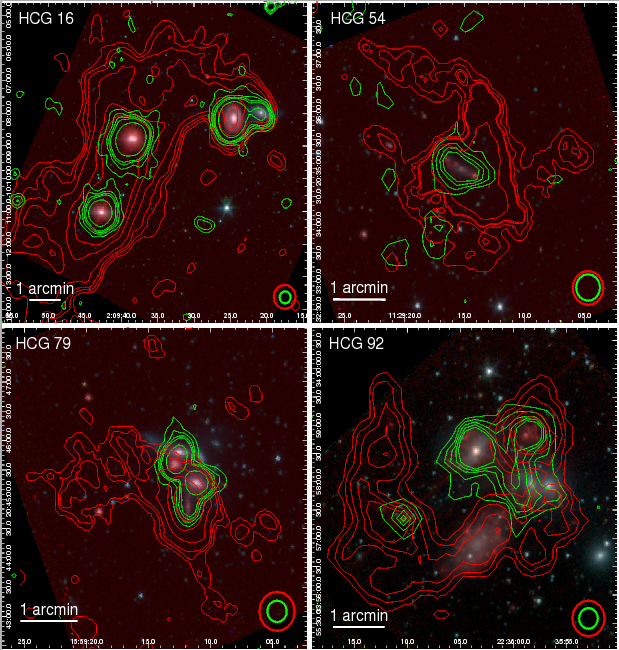}
\caption{True colour images, created using the 3.6, 4.5, and 8.0µm {\it Spitzer}/IRAC mosaics of groups HCG16, HCG54, HCG79, and HCG92, respectively. The green and red contour maps are the 250$\mu$m {\it Herschel}/SPIRE and the {\it VLA}/HI moment-0 maps of the same groups. The five contours correspond to 3, 5, 7, 9, and 20$\sigma$. In the bottom right corner of each panel, we present the size of the SPIRE 250$\mu$m and {\it VLA} band beams, respectively. The R.A. and Dec. scales of the axes are in J2000.0 coordinates.}
\label{fig:fig12}
\end{center}
\end{figure*}

In the far-IR and sub-mm images of HCG16, HCG54, HCG79, and HCG92, we noticed diffuse emission located in the space between the galaxies. Nevertheless, these could be background sources or artefacts created from the detectors and/or the reduction process. To investigate these possibilities, we first convolved all the {\it Herschel}/PACS and SPIRE images to the beam of SPIRE 500$\mu$m. Mapping artefacts should not appear systematically at every {\it Herschel} band. In addition, we compared the remaining sources with our {\it Spitzer}/IRAC 3.6$\mu$m images and the SDSS database and excluded the background sources. In Fig.~\ref{fig:fig12} we present the {\it Spitzer}/IRAC ``true'' colour image of our groups, overlaid on the HI moment-0 (in red; for HCG16, HCG79, and HCG92 taken from \citet{Verdes01} and for HCG54 from \citet{Verdes02}). The {\it Herschel}/SPIRE 250$\mu$m emission is also shown in green. In HCG16 extended dust envelopes are noticeable surrounding the galaxies HCG16a and HCG16b. In the remaining three groups, all galaxies are located in common dust envelopes. In addition, in HCG54 and HCG92, several ``dust clumps'' are found in the proximity of the group, and the compact dust structure located in the south-east of HCG92c also coincides with the location of a peak in the emission of atomic gas. What it is worth noticing here is absence of an one-to-one correlation between the atomic gas and the cold dust; for example, in HCG92c and HCG79b the dynamical interactions had swiped the interstellar gas out of their disks, but these galaxies still have extended cold dust disks. 

The absence of dust in the regions where we detect large HI intragroup clouds could be attributed to the fact that dust emission is below the detection limit. Using the HI moment-0 maps from Verdes-Montenegro et al. (2001, 2002) and the following formula,
\begin{equation}
M_{HI}[M\odot] = 2.35\times10^{5}D^{2}[Mpc]\int S(v) dv [Jy~km~sec^{-1}], 
\end{equation}
we estimate a lower limit in the mass of the diffuse atomic gas of $\sim$10$^{6}$M$_{\odot}$/beam. Assuming a constant atomic gas-to-dust ratio of 150 \citep{Draine07}, we estimate the mass of the diffuse dust in that region to be approximately 7$\times$10$^{4}$M$_{\odot}$. In addition, using the SPIRE 250$\mu$m images, we estimated a typical 3$\sigma$ limit at $\sim$0.067Jy/beam, which corresponds to a lower limit in the mass of the diffuse dust of 12.5$\times$10$^{4}$ M$_{\odot}$/beam (assuming $\beta$=1.5 and T=15K). Therefore, we can rule out the possibility that the diffuse dust emission is below the detection limit of our observations. We conclusively detect cold intragroup dust in at least four HCGs, for which we have atomic gas imaging, but the spatial distribution of the cold dust does not correlate with that of the atomic gas. 

\section{Conclusions}
In this paper we have presented our results of the first far-IR and sub-mm analysis of Hickson compact group galaxies, based on {\it Herschel} observations. We fitted the UV-to-far-IR SEDs and accurately estimated for first time the mass and temperature of the dust for a sample of 120 galaxies, contained in 28 HCGs, as well as a control sample of field galaxies, the KINGFISH sample. Based on the analysis of these results we conclude the following:

\begin{itemize}

\item Late-type galaxies that were found to display red {\it [NUV-r]} colours, owing to the enhanced stellar populations they contain, also have significantly lower dust-to-stellar mass ratios than normal HCG or field ``blue cloud'' late-type galaxies of the same stellar mass. Examining them in more detail, we find that their atomic gas-to-dust mass ratios are more than an order of magnitude lower than the blue cloud late-type galaxies, even though they have similar molecular gas-to-dust mass ratios. We suggest that significant amounts of dust were removed from the disks of these galaxies as a result of tidal and hydrodynamic interactions.

\item Unlike to the dynamically ``young'' groups, where no early-type galaxy member was detected, almost 40\% of the early-type members of the dynamically ``old'' groups were detected in all the far-IR and sub-mm bands of {\it Herschel}. The ``dusty'' lenticular galaxies display very similar dust masses to the red late-type galaxies. Based on the stellar masses, the star formation rates, the {\it [NUV-r]} and mid-IR colours, and the dust-to-stellar and gas mass ratios of the red late-types and the dusty lenticulars, we suggest that they represent transition populations between blue star-forming disk galaxies and quiescent early-type ellipticals in groups. Moreover, MOHEGs seem to dominate this phase of transition.

\item We detect seven ``dusty'' elliptical galaxies, which display high dust-to-stellar mass ratios, comparable to or higher than those of the red late-type galaxies. Unlike to the rest of the sample, there is no correlation between their dust and stellar masses. Moreover, these galaxies display significantly higher sSFRs than the remaining ellipticals in HCGs. We suggest that in these enhanced systems, their dust content is due to accretion from other group members and the intragroup medium, as well as to active star formation.

\item Examining the spatial distribution of the emission in the red late-type galaxies, we find that $\sim$20\% of their 250$\mu$m emission is located outside their main bodies. Similarly, nearly half of the dusty lenticulars also display such extended emission, with the remaining having dust disks that are much smaller than their stellar ones. Estimating the mass of the extended dust we find that it is not enough to compensate for the very low dust-to-stellar mass ratios of these galaxies. 

\item Searching for the missing dust, we conclusively detect cold intragroup dust in at least four HCGs. Depending on the assumptions we make for its temperature and emissivity, we estimate its mass to be approximately 7-76\% of the total dust mass of the group. Comparing the atomic gas and the dust maps and examining the detection limits, we also conclude that there is no correlation in the distributions of the intragroup gas and the cold dust.

\end{itemize}

\begin{acknowledgements}  
TB and VC would like to acknowledge partial support from the EU FP7 Grant PIRSES-GA-2012-316788. This work is based, in part, on observations made with Herschel, a European Space Agency Cornerstone Mission with significant participation by NASA. Support for this work was also provided by NASA through an award issued by JPL/Caltech. We also appreciate the very useful comments of the referee, A. Boselli, which helped improve this paper. Finally, we are grateful to L. Verdes-Montenegro for providing the HI moment-0 maps of 4 HCGs. 
\end{acknowledgements} 



\pagebreak
\clearpage
\onecolumn

\begin{table}[htdp]
\caption{Observational parameters of our sample}
\begin{center}
\begin{tabular}{lccccl}
\hline \hline
Group-ID & PACS channels & Duration$^{\dagger}$ (sec) & SPIRE channels & Duration$^{\dagger}$ (sec) & Proposal PI\\
\hline 
HCG02 & 70/160 & 714 &250/350/500 & 1974 & V. Charmandaris \\
HCG04 & 70/160 & 548 & 250/350/500 & 859 & V. Charmandaris \\
HCG06 & 110/160 & 1106 & 250/350/500 & 445 & P. Appleton \\
HCG07 & 70/160 & 726 & 250/350/500 & 1926 & V. Charmandaris \\
HCG15 & 70/110/160 & 832/906/1738 & 250/350/500 & 1541 & V. Charmandaris, P. Appleton \\
HCG16 & 70/110/160 & 606/606/1212 & 250/350/500 & 1324 & D. Sanders* \\
HCG22 & 70/160 & 832 & 250/350/500 & 859 & V.Charmandaris \\
HCG25 & 110/160 & 630 & 250/350/500 & 1541 & P. Appleton \\
HCG37 & 70/160 & 684 & 250/350/500 & 859 & V. Charmandaris \\
HCG38 & 70/160 & 684 & 250/350/500 & 859 & V. Charmandaris \\
HCG40 & 70/110/160 & 548/516/1064 & 250/350/500 & 445 & V. Charmandaris, P. Appleton \\
HCG44 & 70/160 & 2708 & 250/350/500 & 1035 & R. Kennicutt* \\
HCG47 & 70/160 & 684 & 250/350/500 & 859 & V. Charmandaris \\
HCG54 & 70/160 & 512 & 250/350/500 & 859 & V. Charmandaris \\
HCG55 & 70/160 & 394 & 250/350/500 & 445 & V. Charmandaris, M. Cluver \\
HCG56 & 70/160 & 466 & 250/350/500 & 445 & V. Charmandaris, M. Cluver \\
HCG57 & 110/160 & 1068 & 250/350/500 & 999 & P. Appleton, M. Cluver \\
HCG59 & 70/160 & 868 & 250/350/500 & 859 & V. Charmandaris \\
HCG68 & 70/160 & 1480 & 250/350/500 & 1654 & V. Charmandaris, M. Cluver \\
HCG71 & 70/160 & 832 & 250/350/500 & 859 & V. Charmandaris \\
HCG75 & 70/160 & 548 & 250/350/500 & 859 & V. Charmandaris \\
HCG79 & 70/160 & 424 & 250/350/500 & 445 & V. Charmandaris, M. Cluver \\
HCG82 & 70/160 & 684 & 250/350/500 & 859 & V. Charmandaris \\
HCG91 & 70/160 & 1480 & 250/350/500 & 999 & V. Charmandaris, M. Cluver\\
HCG92 & 110/160 & 48600 & 250/350/500 & 583 & P. Guillard* \\
HCG95 & 110/160 & 758 & 250/350/500 & 445 & P. Appleton \\
HCG96 & 70/110/160 & 396/396/792 & 250/350/500 & 445 & D. Sanders*, M. Cluver \\
HCG100 & 70/160 & 684 & 250/350/500 & 445 & V. Charmandaris, M. Cluver \\
\hline
\end{tabular}
\end{center}
\label{tab:observations}
$^{\dagger}$ On-source time.\\
** Obtained from the {\it Herschel} Science Archive (HSA)\\
\end{table}%

\begin{table}[htdp]
\caption{Sample of the far-IR and sub-mm photometry.}
\begin{center}
\begin{tabular}{ccccccccccccc}
\hline\hline
Galaxy  & PACS 70$\mu$m$^{a}$ & PACS 110$\mu$m$^{a}$ & PACS 160$\mu$m$^{a}$ & SPIRE 250$\mu$m & SPIRE 350$\mu$m & SPIRE 500$\mu$m \\
      ID     & (Jy)            & (Jy)             & (Jy)             & (Jy)              & (Jy)              & (Jy)              \\
\hline
  HCG16a & 0.413$\pm$0.021 & 0.697$\pm$0.035 & 1.026$\pm$0.051 & 0.503$\pm$0.109 & 0.350$\pm$0.057 & $<$0.040$\pm$ 0.030 \\
  HCG16b & 6.836$\pm$0.342 & 10.95$\pm$0.55   & 10.71$\pm$0.54   & 4.271$\pm$0.331 & 1.558$\pm$0.130 & 0.485$\pm$0.052 \\
  HCG16c & 13.30$\pm$0.66   & 15.99$\pm$0.80    & 12.61$\pm$0.63   & 3.984$\pm$0.291 & 1.379$\pm$0.106 & 0.385$\pm$.037 \\
  HCG16d & 12.04$\pm$0.60   & 12.21$\pm$0.61   & 8.234$\pm$0.411 & 2.222$\pm$0.190 & 0.737$\pm$0.075 & 0.210$\pm$.035 \\
  HCG40a & $<$0.249$\pm$0.012$^{b}$ & 0.103$\pm$0.005 & 0.195$\pm$0.010 & $<$0.040$\pm$0.030 & $<$0.026$\pm$0.020 & $<$0.045$\pm$0.030 \\
  HCG40b & 0.121$\pm$0.006 & 0.306$\pm$0.015& 0.318$\pm$0.016 & $<$0.030$\pm$0.020 & $<$0.026$\pm$0.012 & $<$0.045$\pm$0.009 \\
  HCG40c & 0.991$\pm$0.049 & 2.258$\pm$0.113 & 3.002$\pm$0.150 & 1.260$\pm$0.096 & 0.580$\pm$0.046 & 0.230$\pm$0.021 \\
  HCG40d & 1.103$\pm$0.055 & 1.710$\pm$0.086 & 1.625$\pm$0.081 & 0.640$\pm$0.051 & 0.280$\pm$0.025 & 0.100$\pm$0.013 \\
  HCG40e& 0.072$\pm$0.004 & 0.134$\pm$0.007 & 0.368$\pm$0.018 & $<$0.030$\pm$0.022 & $<$0.026$\pm$0.013 & $<$0.045$\pm$0.010 \\
\hline
\end{tabular}
\end{center}
\label{tab:fluxes}
$^{a}$ The PACS photometry presented here was performed on the maps reduced by the Scanamoprhos mapper.\\
$^{b}$ With upper limits are marked objects whose emission was under the detection limit.\\
The full table is available at the CDS.
\end{table}%

\begin{table}[htdp]
\caption{Derived physical parameters of our galaxies based on the SED modelling}
\begin{center}
\begin{tabular}{lcccccccc}
\hline \hline
  Galaxy & Morphology$^{a}$ & Group type$^{b}$ & log sSFR  & SFR  & log M$_{star}$  & log L$_{IR}$ & log M$_{dust}$ & T$_{dust}^{c}$\\
     &  &  & (yr$^{-1}$)  & (M$_{\odot}$ yr$^{-1}$)  & (M$_{\odot}$)  & (L$_{\odot}$) & (M$_{\odot}$) & (K) \\
\hline
  HCG02a & LT (3)& DY & -9.47  &  2.35 & 9.86  & 10.25 & 7.24 & 32.1 \\
  HCG02b & LT (9)& DY & -9.27  &  2.83 & 9.72  & 10.47 & 6.80 & 34.9 \\
  HCG02c & LT (5)& DY & -10.02 &  0.48 & 9.72  & 9.62  & 7.01 & 27.8 \\
  HCG04a & LT (4)& DO & -9.42  &  16.4 & 10.63 & 11.22 & 7.80 & 31.3 \\
  HCG04b & LT (3)& DO & -10.22 &  0.54 & 9.96  & 9.74  & 7.32 & 29.1 \\
  HCG04d & ET (-5)& DO & -10.17 &  0.70 & 10.04 & 10.01 & 7.16 & 41.9 \\
  HCG06a & ET (0)& DO & -12.22 &  0.04 & 10.90 & $<$9.35$^{d}$  & $<$6.77$^{d}$ & $<$48.4 \\
  HCG06b & LT (2)& DO & -12.06 &  0.07 & 10.86 & 9.80  & 7.02 & 51.6 \\
  HCG06c & ET (-5)& DO & -12.31 &  0.02 & 10.68 & $<$9.16  & $<$6.76 & $<$50.2 \\
  HCG06d & LT (4)& DO & -10.86 &  0.08 & 9.77  & 8.73  & 6.07 & 28.9 \\
  HCG07a & LT (2)& DY & -10.67 &  2.09 & 10.94 & 10.60 & 7.37 & 45.1 \\
  HCG07b & ET (-2)& DY & -11.31 &  0.17 & 10.56 & $<$9.13  & $<$6.34 & $<$22.7 \\
  HCG07c & LT (2)& DY & -10.22 &  1.55 & 10.40 & 10.11 & 7.42 & 42.3 \\
  HCG07d & LT (2)& DY & -10.56 &  0.28 & 10.02 & 9.31  & 6.67 & 28.0 \\
  HCG15a & ET (-2)& DO & -11.92 &  0.10 & 10.97 & 9.61  & 7.27 & 45.6 \\
  HCG15b & ET (-3)& DO & -12.36 &  0.02 & 10.77 & $<$9.41  & $<$6.40 & $<$51.7 \\
  HCG15c & ET (-3)& DO & -12.97 &  0.01 & 11.11 & $<$9.43  & $<$6.37 & $<$52.2 \\
  HCG15d & ET (-3)& DO & -11.27 &  0.07 & 10.11 & 8.86  & 6.30 & 34.9 \\
  HCG15e & LT (1)& DO & -11.56 &  0.06 & 10.39 & 8.83  & 6.45 & 35.4 \\
  HCG15f & LT (4)& DO & -9.875 &  0.33 & 9.37  & 9.32  & 6.77 & 23.5 \\
  HCG16a & LT (2)& DY & -10.92 &  1.05 & 10.92 & 10.02 & 6.94 & 35.8 \\
  HCG16b & LT (1)& DY & -12.22 &  0.05 & 10.97 & 10.47 & 7.39 & 43.8 \\
  HCG16c & LT (10)& DY & -9.17  &  11.5 & 10.25 & 11.09 & 7.30 & 33.2 \\
  HCG16d & LT (10)& DY & -10.27 &  1.16 & 10.35 & 10.94 & 7.02 & 31.8 \\
  HCG22a & ET (-5)& DO & -12.36 &  0.02 & 10.72 & 8.80  & 5.45 & 47.8 \\
  HCG22b & LT (-2)& DO & -11.52 &  0.01 & 9.64  & $<$8.25  & $<$4.92 & $<$34.5 \\
  HCG22c & LT (8)& DO & -10.22 &  0.19 & 9.50  & 9.02  & 6.78 & 37.1 \\
  HCG25a & LT (5)& DO & -9.82  &  2.40 & 10.22 & 10.39 & 7.52 & 26.3 \\
  HCG25b & LT (1)& DO & -11.67 &  0.16 & 10.89 & 9.62  & 6.77 & 42.3 \\
  HCG25d & ET (-2)& DO & -11.17 &  0.09 & 10.12 & $<$8.95  & $<$5.35 & $<$29.9 \\
  HCG25f & ET (-2)& DO & -10.92 &  0.07 & 9.84  & $<$8.73  & $<$5.09 & $<$20.8 \\
  HCG37a & ET (-5)& DO & -11.17 &  1.46 & 11.32 & $<$11.10 & $<$7.99 & $<$35.6 \\
  HCG37b & LT (3)& DO & -11.22 &  0.52 & 10.93 & 10.36 & 7.27 & 50.6 \\
  HCG37c & ET (0)& DO & -11.47 &  0.05 & 10.19 & 9.10  & 6.79 & 41.1 \\
  HCG37d & LT (8)& DO & -10.86 &  0.02 & 9.32  & 9.21  & 6.73 & 45.4 \\
  HCG37e & ET (-5)& DO & -10.86 &  0.04 & 9.50  & 8.86  & 6.45 & 36.3 \\
  HCG38a & LT (4)& DY & -10.56 &  1.78 & 10.80 & 10.54 & 7.52 & 39.9 \\
  HCG38b & LT (7)& DY & -10.52 &  1.66 & 10.73 & 10.59 & 7.59 & 39.7 \\
  HCG38c & LT (10)& DY & -10.47 &  1.02 & 10.51 & 10.26 & 7.78 & 37.5 \\
  HCG40a & ET (-5)& DO & -11.86 &  0.13 & 10.97 & 9.31  & 7.04 & 37.3 \\
  HCG40b & ET (0)& DO & -12.06 &  0.05 & 10.80 & 9.56  & 6.72 & 42.1 \\
  HCG40c & LT (3)& DO & -11.11 &  0.67 & 10.97 & 10.48 & 7.56 & 47.2 \\
  HCG40d & LT (1)& DO & -10.86 &  0.65 & 10.67 & 10.43 & 7.11 & 49.0 \\
  HCG40e & LT (1)& DO & -11.27 &  0.09 & 10.27 & 9.53  & 7.47 & 41.0 \\
\hline
\end{tabular}
\end{center}
\label{tab:properties}
\end{table}%
 
 \begin{table*}
\begin{center}
\begin{tabular}{lcccccccc}
\hline \hline
  Galaxy & Morphology$^{a}$ & Group Type$^{b}$ & log sSFR  & SFR  & log M$_{star}$  & log L$_{IR}$ & log M$_{dust}$ & T$_{dust}^{c}$\\
     &  &  & (yr$^{-1}$)  & (M$_{\odot}$ yr$^{-1}$)  & (M$_{\odot}$)  & (L$_{\odot}$) & (M$_{\odot}$) & (K) \\
\hline
  HCG44a & LT (1)& DY & -11.67 &  0.07 & 10.55 & 9.78  & 6.84 & 41.9 \\
  HCG44b & ET (-5)& DY & -12.27 &  0.01 & 10.31 & $<$8.47  & $<$5.21 & $<$40.1 \\
  HCG44c & LT (1)& DY & -10.06 &  0.15 & 9.24  & 9.35  & 6.21 & 39.3 \\
  HCG44d & LT (5)& DY & -9.32  &  0.28 & 8.75  & 9.25  & 6.50 & 24.7 \\
  HCG47a & LT (3)& DY & -10.56 &  3.44 & 11.11 & 10.77 & 7.60 & 43.6 \\
  HCG47b & ET (-2)& DY & -11.36 &  0.20 & 10.68 & $<$9.06  & $<$6.53 & $<$22.3 \\
  HCG47c & LT (3)& DY & -10.67 &  0.28 & 10.11 & 9.39  & 6.26 & 21.9 \\
  HCG47d & LT (2)& DY & -10.52 &  0.53 & 10.26 & 9.36  & 5.60 & 32.3 \\
  HCG54a & LT (8)& DY & -10.77 &  0.01 & 8.26  & 8.02  & 4.71 & 36.4 \\
  HCG54b & LT (10)& DY & -9.97  &  0.01 & 8.25  & 8.28  & 5.22 & 33.5 \\
  HCG54c & LT (10)& DY & -9.47  &  0.01 & 7.59  & 7.86  & 5.49 & 28.7 \\
  HCG54d & LT (10)& DY & -9.67  &  0.01 & 7.32  & 7.35  & 4.58 & 30.6 \\
  HCG55a & ET (0)& DO & -11.56 &  0.27 & 11.00 & $<$9.44  & $<$6.02 & $<$33.0 \\
  HCG55b & ET (-5)& DO & -12.42 &  0.03 & 10.92 & 10.14 & 7.58 & 49.7 \\
  HCG55c & ET (-5)& DO & -10.92 &  1.32 & 11.06 & $<$10.79 & $<$8.18 & $<$41.8 \\
  HCG55d & ET (-2)& DO & -11.31 &  0.16 & 10.55 & $<$9.24  & $<$6.15 & $<$42.2 \\
  HCG56a & LT (5)& DO & -10.67 &  0.57 & 10.42 & 9.97  & 7.22 & 45.6 \\
  HCG56b$^{\dagger}$ & ET (-2)& DO & -11.56 &  0.13 & 10.67 & 10.08 & 6.70 & 45.4 \\
  HCG56c & ET (0)& DO & -12.36 &  0.01 & 10.64 & $<$9.05  & $<$6.14 & $<$36.5 \\
  HCG56d & ET (0)& DO & -10.56 &  0.39 & 10.17 & 10.07 & 6.80 & 47.2 \\
  HCG56e & ET (-2)& DO & -10.22 &  0.24 & 9.62  & 9.46  & 6.91 & 28.5 \\
  HCG57a & LT (2)& DO & -11.81 &  0.32 & 11.31 & 10.24 & 7.60 & 48.5 \\
  HCG57b & LT (3)& DO & -11.42 &  0.37 & 11.01 & 9.79  & 7.45 & 49.2 \\
  HCG57c & ET (0)& DO & -12.97 &  0.01 & 10.94 & $<$9.27$^{d}$  & $<$6.19$^{d}$ & 42.4 \\
  HCG57d & LT (3)& DO & -10.11 &  1.74 & 10.39 & 10.18 & 6.89 & 39.4 \\
  HCG57e & ET (-2)& DO & -11.97 &  0.04 & 10.59 & 9.52  & 6.94 & 49.9 \\
  HCG57f & ET (-3)& DO & -12.97 &  0.01 & 10.76 & $<$9.07  & $<$6.21 & 51.8 \\
  HCG57g & ET (-3)& DO & -11.86 &  0.06 & 10.68 & $<$8.84  & $<$6.62 & 22.3 \\
  HCG57h & LT (3)& DO & -11.06 &  0.10 & 10.10 & 9.19  & 6.32 & 44.8 \\
  HCG59a & ET (-5)& DY & -12.97 &  0.01 & 10.38 & 9.78  & 6.46 & 54.7 \\
  HCG59b & ET (-2)& DY & -11.97 &  0.01 & 9.88  & $<$8.30  & $<$5.07 & 42.8 \\
  HCG59c & LT (5)& DY & -10.27 &  0.10 & 9.24  & 8.95  & 6.38 & 25.4 \\
  HCG59d & LT (10)& DY & -9.87  &  0.05 & 8.66  & 8.96  & 6.14 & 41.4 \\
  HCG68a & ET (-2)& DO & -10.97 &  0.39 & 10.56 & 8.94  & 5.97 & 37.4 \\
  HCG68b & ET (-2)& DO & -11.92 &  0.06 & 10.69 & 9.22  & 6.48 & 48.6 \\
  HCG68c & LT (3)& DO & -12.97 &  0.01 & 10.93 & 10.14 & 7.23 & 54.3 \\
  HCG68d & ET (-2)& DO & -12.27 &  0.01 & 9.96  & 8.11  & 5.18 & 45.9 \\
  HCG68e & ET (0)& DO & -11.56 &  0.01 & 9.67  & $<$7.68  & $<$5.21 & 36.5 \\
  HCG71a & LT (6)& DY & -10.56 &  2.05 & 10.89 & 10.35 & 7.70 & 42.7 \\
  HCG71b & LT (3)& DY & -10.22 &  2.09 & 10.56 & 10.64 & 7.45 & 40.4 \\
  HCG71c & LT (3)& DY & -10.02 &  0.49 & 9.71  & 9.49  & 6.95 & 24.4 \\
  HCG75a & ET (-5)& DO & -11.61 &  0.36 & 11.17 & $<$9.36  & $<$5.67 & 27.9 \\
  HCG75b & LT (3)& DO & -12.52 &  0.01 & 10.54 & $<$8.71  & $<$5.60 & 35.6 \\
  HCG75c & ET (-2)& DO & -11.86 &  0.08 & 10.77 & 9.60  & 6.83 & 40.8 \\
  HCG75d & LT (7)& DO & -10.56 &  1.01 & 10.57 & 10.22 & 7.66 & 35.2 \\
  HCG75e & LT (1)& DO & -11.47 &  0.11 & 10.55 & $<$9.04  & $<$5.51 & 28.2 \\
\hline
\end{tabular}
\end{center}
\label{tab:properties}
\end{table*}%
 
\begin{table*}
\begin{center}
\begin{tabular}{lcccccccc}
\hline \hline
  Galaxy & Morphology$^{a}$ & Group Type$^{b}$ & log sSFR  & SFR  & log M$_{star}$  & log L$_{IR}$ & log M$_{dust}$ & T$_{dust}^{c}$\\
     &  &  & (yr$^{-1}$)  & (M$_{\odot}$ yr$^{-1}$)  & (M$_{\odot}$)  & (L$_{\odot}$) & (M$_{\odot}$) & (K) \\
\hline
  HCG75f & ET (-2)& DO & -12.11 &  0.04 & 10.72 & 9.89  & 7.11 & 52.7 \\
  HCG79a & ET (5)& DO & -11.56 &  0.07 & 10.46 & 9.80  & 6.63 & 50.5 \\
  HCG79b & ET (-2)& DO & -11.47 &  0.06 & 10.27 & 9.59  & 6.34 & 51.4 \\
  HCG79c & ET (-2)& DO & -12.06 &  0.01 & 9.78  & 8.48  & 5.25 & 47.5 \\
  HCG79d & LT (5)& DO & -10.36 &  0.06 & 9.03  & 8.95  & 6.43 & 41.4 \\
  HCG82a & ET (-2)& DO & -12.36 &  0.08 & 11.30 & $<$9.72  & $<$6.59 & 39.1 \\
  HCG82b & ET (-2)& DO & -12.11 &  0.09 & 11.10 & 9.94  & 7.04 & 51.5 \\
  HCG82c & LT (10)& DO & -9.97  &  4.09 & 10.57 & 10.83 & 7.55 & 40.1 \\
  HCG82d & ET (0)& DO & -12.27 &  0.01 & 10.48 & $<$8.76  & $<$6.83 & 46.3 \\
  HCG91a & LT (4)& DY & -9.97  &  11.8 & 11.06 & 11.06 & 7.84 & 39.9 \\
  HCG91b & LT (5)& DY & -9.97  &  4.75 & 10.60 & 10.81 & 7.50 & 36.6 \\
  HCG91c & LT (3)& DY & -9.92  &  2.19 & 10.27 & 10.09 & 7.46 & 17.5 \\
  HCG91d & ET (-2)& DY & -10.97 &  0.32 & 10.50 & $<$9.29  & $<$5.59 & 27.6 \\
  HCG92b & LT (4)& DO & -11.17 &  0.52 & 10.89 & 9.34  & 6.97 & 31.3 \\
  HCG92c & LT (4)& DO & -12.12 &  0.08 & 11.06 & 10.27 & 7.27 & 43.1 \\
  HCG92d & ET (-5)& DO &-12.22  &  0.05 & 10.97 & $<$8.40  & $<$6.33 & 36.5 \\
  HCG92e & ET (-5)& DO & -12.22 &  0.03 & 10.82 & $<$8.27  & $<$6.27 & 26.9 \\
  NGC7320c & ET (0)& DO &-12.02&  0.01 & 10.07 & 8.94  & 6.03 & 43.4 \\
  HCG95a & ET (-5)& DY & -11.36 &  0.44 & 11.01 & $<$9.56  & $<$6.28 & 32.9 \\
  HCG95b & LT (6)& DY & -10.11 &  3.73 & 10.71 & 10.77 & 7.72 & 30.9 \\
  HCG95c & LT (9)& DY & -10.02 &  2.40 & 10.47 & 10.51 & 7.65 & 30.7 \\
  HCG95d & LT (3)& DY & -11.11 &  0.46 & 10.79 & 10.09 & 7.43 & 43.8 \\
  HCG96a & LT (4)& DY & -10.06 &  17.4 & 11.31 & 11.38 & 7.82 & 40.5 \\
  HCG96b & ET (-3)& DY & -11.67 &  0.15 & 10.89 & $<$9.55  & $<$6.34 & 38.4 \\
  HCG96c & LT (1)& DY & -10.77 &  0.94 & 10.72 & 10.37 & 6.99 & 39.5 \\
  HCG96d & LT (10)& DY & -9.32  &  0.68 & 9.13  & 9.71  & 6.32 & 25.1 \\
  HCG100a & LT (0)& DY & -10.31&  2.05 & 10.60 & 10.60 & 7.11 & 45.6 \\
  HCG100b & LT (1)& DY & -9.67 &  0.48 & 9.33  & 9.88  & 6.72 & 39.5 \\
  HCG100c & LT (3)& DY & -9.77 &  0.92 & 9.73  & 9.52  & 6.59 & 42.1 \\
  HCG100d & LT (6)& DY & -11.42&  0.01 & 9.06  & 9.00  & 6.11 & 53.0 \\
\hline
\end{tabular}
\end{center}
$^{a}$ With LT we indicate the late-type galaxies -- spirals (1$\le$T$\le$7) and irregulars (T$\ge$8) -- and with ET the early-type ones -- ellipticals (T$\le$-4) and lenticulars (-3$\le$T$\le$0).\\
$^{b}$ Group dynamical state (DY: dynamically young, DO: dynamically old)\\
$^{c}$ Temperature of the dust in the ambient ISM and the stellar birth clouds (T$_{dust}$=f$_{\mu}$T$_{cold}^{ISM}$+[1-f$_{\mu}$]T$_{warm}^{BC}$, where f$_{\mu}$ is the fraction of the total dust luminosity contributed from the ambient ISM).\\
$^{d}$ Upper limits mark the dust masses and luminosities of the galaxies not detected in the {\it Herschel} bands\\
$^{\dagger}$ Galaxy HCG56b is flagged because it hosts a powerful AGN that dominates its mid-IR colours.

\end{table*}%

\begin{table}[htdp]
\caption{Derived physical parameters of the galaxies in the KINGFISH sample based on the SED modelling}
\begin{center}
\begin{tabular}{lccccccc}
\hline \hline
  Galaxy & Morphology* & log sSFR  & SFR  & log M$_{star}$  & log L$_{IR}$ & log M$_{dust}$ & T$_{dust}$\\
     &  & (yr$^{-1}$)  & (M$_{\odot}$ yr$^{-1}$)  & (M$_{\odot}$)  & (L$_{\odot}$) & (M$_{\odot}$) & (K) \\
\hline
  NGC0337 & LT &  -9.62 &  1.62 &  9.77 & 10.27 & 7.26 & 35.2 \\
  NGC0628 & LT & -10.02 &  1.20 & 10.09 & 10.10 & 7.50 & 26.3 \\
  NGC0855 & LT & -10.06 &  0.03 &  8.59 &  8.47 & 5.79 & 34.7 \\
  NGC0925 & LT & -10.06 &  0.30 &  9.53 &  9.43 & 7.16 & 25.7 \\
  NGC1097 & LT & -10.17 &  5.88 & 10.92 & 10.81 & 7.94 & 30.9 \\
  NGC1266 & ET & -10.27 &  0.46 & 9.95 & 10.54 & 6.70 & 34.9 \\
  NGC1291 & LT & -11.42 &  0.23 & 10.80 &  9.52 & 7.17 & 32.7 \\
  NGC1316 & LT & -12.97 &  0.04 & 11.64 & 10.10 & 6.95 & 27.0 \\
  NGC1377 & ET & -10.02 &  0.85 &  9.96 & 10.25 & 6.26 & 55.7 \\
  IC0342  & LT & -10.67 &  0.01 &  8.61 &  8.46 & 5.67 & 30.9 \\
  NGC1482 & LT & -10.31 &  1.28 & 10.43 & 10.89 & 7.41 & 39.7 \\
  NGC1512 & LT & -10.61 &  0.42 & 10.18 &  9.56 & 7.36 & 24.7 \\
  NGC2146 & LT & -10.31 &  1.34 & 10.44 & 10.90 & 7.32 & 41.2 \\
  HoII    & LT &  -9.37 &  0.01 &  7.55 &  7.53 & 4.43 & 44.5 \\
  DDO053  & LT &  -9.32 &  0.01 &  7.00 &  7.00 & 3.00 & 42.0 \\
  NGC2798 & LT &  -9.52 &  2.81 &  9.96 & 10.56 & 7.00 & 43.2 \\
  NGC2841 & LT & -11.11 &  0.24 & 10.51 &  9.62 & 7.25 & 23.2 \\
  NGC2915 & LT & -10.52 &  0.01 &  8.68 &  8.03 & 5.25 & 31.9 \\
  HoI     & LT &  -9.82 &  0.01 &  7.00 &  7.00 & 4.01 & 31.1 \\
  NGC2976 & LT & -10.52 &  0.01 &  7.00 &  7.00 & 3.00 & 25.9 \\
  NGC3049 & LT &  -9.72 &  0.56 &  9.44 &  9.65 & 6.79 & 45.1 \\
  NGC3077 & LT & -10.97 &  0.01 &  7.00 &  7.00 & 3.18 & 35.9 \\
  M81dwB  & LT & -10.31 &  0.01 &  7.38 &  7.00 & 4.66 & 35.9 \\
  NGC3190 & LT & -11.86 &  0.03 & 10.47 &  9.77 & 6.98 & 25.9 \\
  NGC3184 & LT & -10.17 &  0.53 &  9.88 &  9.67 & 7.11 & 30.6 \\
  NGC3198 & LT & -10.22 &  0.38 &  9.81 &  9.56 & 7.03 & 34.9 \\
  IC2574  & LT &  -9.82 &  0.01 &  7.27 &  7.00 & 4.63 & 37.4 \\
  NGC3265 & ET & -10.17 &  0.13 &  9.28 &  9.38 & 6.13 & 41.6 \\
  NGC3351 & LT & -10.56 &  0.69 & 10.42 &  9.97 & 7.19 & 29.3 \\
  NGC3521 & LT & -10.61 &  1.62 & 10.82 & 10.59 & 7.74 & 24.9 \\
  NGC3621 & LT & -10.81 &  0.48 & 10.48 & 10.35 & 7.57 & 43.9 \\
  NGC3627 & LT & -10.22 &  1.81 & 10.52 & 10.52 & 7.45 & 31.1 \\
  NGC3373 & LT &  -8.97 &  0.25 &  8.35 &  8.88 & 5.82 & 42.6 \\
  NGC3938 & LT &  -9.92 &  0.64 &  9.69 &  9.87 & 7.15 & 27.7 \\
  NGC4236 & LT & -10.22 &  0.01 &  7.00 &  7.00 & 3.00 & 34.8 \\
  NGC4254 & LT &  -9.67 &  21.3 & 11.18 & 11.34 & 8.39 & 31.9 \\
  NGC4321 & LT & -10.27 &  5.49 & 11.01 & 10.89 & 8.11 & 29.1 \\
  NGC4536 & LT &  -9.82 &  6.30 & 10.63 & 10.85 & 7.88 & 38.1 \\
  NGC4559 & LT &  -9.97 &  0.97 &  9.88 &  9.85 & 7.34 & 28.8 \\
  NGC4569 & LT & -10.72 &  0.02 &  9.07 &  8.79 & 5.97 & 30.1 \\
  NGC4579 & LT & -11.36 &  0.40 & 10.97 & 10.15 & 7.58 & 22.6 \\
  NGC4625 & LT &  -9.92 &  0.07 &  8.82 &  8.62 & 6.11 & 26.0 \\
  NGC4631 & LT &  -9.62 &  2.63 & 10.02 & 10.47 & 7.59 & 29.9 \\
  NGC4725 & LT & -10.92 &  1.17 & 10.98 & 10.14 & 7.84 & 24.4 \\
  NGC4736 & LT & -10.67 &  0.38 & 10.23 &  9.57 & 6.57 & 27.1 \\
 \hline
\end{tabular}
\end{center}
\label{tab:KINGFISH}
\end{table}%

\begin{table*}[htdp]
\begin{center}
\begin{tabular}{lccccccc}
\hline \hline
  Galaxy & Morphology* & log sSFR  & SFR  & log M$_{star}$  & log L$_{IR}$ & log M$_{dust}$ & T$_{dust}$\\
     &  & (yr$^{-1}$)  & (M$_{\odot}$ yr$^{-1}$)  & (M$_{\odot}$)  & (L$_{\odot}$) & (M$_{\odot}$) & (K) \\
\hline
  NGC4826 & LT & -11.06 &  0.15 & 10.25 &  9.67 & 6.53 & 31.8 \\
  NGC5055 & LT & -10.61 &  0.79 & 10.55 & 10.14 & 7.48 & 26.9 \\
  NGC5398 & LT &  -9.32 &  0.37 &  8.91 &  9.33 & 6.61 & 43.1 \\
  NGC5457 & LT &  -9.72 &  0.75 &  9.59 &  9.71 & 7.11 & 31.2 \\
  NGC5408 & LT &  -9.97 &  0.05 &  8.74 &  8.64 & 5.06 & 52.6 \\
  NGC5474 & LT &  -9.97 &  0.03 &  8.48 &  8.17 & 5.81 & 34.7 \\
  NGC5713 & LT & -10.11 &  2.75 & 10.57 & 10.72 & 7.42 & 37.7 \\
  NGC5866 & ET & -12.06 &  0.01 & 10.10 &  9.26 & 6.28 & 27.3 \\
  NGC6946 & LT &  -9.72 &  0.03 &  8.19 &  8.40 & 5.56 & 28.3 \\
  NGC7331 & LT & -10.72 &  0.95 & 10.71 & 10.32 & 7.59 & 23.5 \\
  NGC7793 & LT &  -9.47 &  0.21 &  8.81 &  9.07 & 6.64 & 25.3 \\
 \hline
\end{tabular}
\end{center}
\label{tab:KINGFISH}
* LT mark the late-type galaxies and ET the early-type ones\\
\end{table*}%

\begin{table}[htdp]
\caption{Diffuse cold dust emission around red late-type (red-LTGs) and the dusty early-type (dusty-ETGs) galaxies, as well as normal HCG late-type galaxies displaying extended dust features.}
\begin{center}
\begin{tabular}{lccccccc}
\hline\hline
Galaxy-ID & Stellar disk$^{a}$ & Dust disk$^{b}$ & Extended flux & Fraction of & M$_{D,def.}$$^{d}$ & Fraction of  & M$_{ExtD}$$^{f}$ \\
 & (arcsec) & (arcsec) & (Jy) & the total flux$^{c}$ & $\times$10$^{6}$ (M$_{\odot}$) & dust deficit$^{e}$ & $\times$10$^{6}$ (M$_{\odot}$) \\
\hline \\
\multicolumn{8}{c}{Red late-type galaxies} \\
\hline
HCG06b & 25 & 66 & 0.018 & 14\% & 19.67$\pm$7.26 & 65\% & 0.28 -- 0.10 \\
HCG15e & 28 & 48 & 0.008 & 18\% & 10.56$\pm$3.90 & 79\% & 0.05 -- 0.02  \\
HCG16b$^{g}$ & 39 & 24 & -- & -- & 11.64$\pm$4.29 & 32\% & -- \\
HCG25b & 34 & 34 & no extend & 0\% & 25.41$\pm$9.38 & 81\% & 0.00 \\
HCG37b & 32 & 42 & 0.080 & 8\% & 14.89$\pm$5.49 & 44\% & 0.44 -- 0.16\\
HCG40d & 22 & 42 & 0.015 & 19\% & 8.51$\pm$3.14 & 39\% & 0.08 -- 0.03\\
HCG40e & 22 & 42 & 0.027 & 10\% & no deficit & 0\% & 0.14 -- 0.05\\
HCG44a & 107 & 107 & no extend & 0\% & 10.57$\pm$3.90 & 60\% & 0.00 \\
HCG57a & 24 & 32 & 0.150 & 16\% & 24.15$\pm$8.91 & 37\% & 1.37 -- 0.49\\
HCG57b & 30 & 40 & 0.088 & 22\% & 9.84$\pm$3.63 & 26\% & 0.86 --0.31\\
HCG57h$^{g}$ & 19 & 16 & -- & -- & 6.05$\pm$2.23 & 74\% & -- \\
HCG68c & 66 & 108 & 1.110 & 15\% & 16.33$\pm$6.03 & 49\% & 0.71 -- 0.26\\
HCG92b & 22 & 55 & 0.834$^{h}$ & 98\% & 22.03$\pm$8.13 & 70\% & 6.10 -- 2.21\\
HCG92c & 36 & 55 & 0.070$^{h}$ & 41\% & 23.88$\pm$8.12 & 56\% & 0.38 -- 0.14\\
HCG95d & 17 & 27 & 0.031 & 15\% & no deficit & 0\% & 0.57 -- 0.20  \\
\hline \\
\multicolumn{8}{c}{Dusty lenticular galaxies} \\
\hline
HCG15a$^{g}$ & 48 & 29 & -- & -- & 66.22$\pm$24.44 & 52\% & -- \\
HCG37c & 12 & 19 & 0.023 & 34\% & 3.20$\pm$1.18 & 34\% & 0.15 -- 0.05 \\
HCG40b & 17 & -- & -- & -- & 16.52$\pm$6.09 & 61\% & -- \\
HCG55b & 17 & 20 & 0.014 & 7\% & no deficit & 0\% & 0.41 -- 0.20 \\
HCG56b$^{g}$$^{\dagger}$ & 23 & 15 & -- & -- & -- & -- & -- \\
HCG56d & 19 & 24 & 0.013 & 9\% & no deficit & 0\% & 0.11 -- 0.04 \\
HCG56e & 15 & 18 & 0.004 & 5\%  & no deficit & 0\% & 0.03 -- 0.01 \\
HCG57e & 22 & 28 & 0.020 & 14\%  & 8.18$\pm$3.02 & 44\%  & 0.19 -- 0.07 \\
HCG68a$^{g}$ & 51 & 34 & -- & -- & 12.41$\pm$4.58 & 68\%  & -- \\
HCG75c-d & 18-20 & 51 & 0.051 & 16\%  & 24.38$\pm$8.99 & 95\%  & 0.92 -- 0.33 \\
HCG75f & 20 & 29 & 0.026 & 18\%  & no deficit & 0\% & 0.53 -- 0.19 \\
HCG79a & 21 & 27 & 0.067 & 16\%  & 12.85$\pm$4.74 & 85\%  & 0.17 -- 0.06 \\
HCG79b & 35 & 37 & 0.026 & 5\%  & no deficit & 0\%& 0.06 -- 0.02 \\
HCG79c & 14 & 14 & no extend & 0\% & 0.06$\pm$0.02 & 2\%  & 0.00 \\
HCG82b$^{g}$ & 24 & 18 & -- & -- & 26.91$\pm$9.93 & 59\%  & -- \\
NGC7320c & 18 & 28 & 0.041 & 19\% & 5.38$\pm$1.98 & 68\% & 0.18 -- 0.06 \\
\hline
\end{tabular}
\end{center}
\label{tab:mass_diffuse_dust}
\end{table}%

\begin{table*}[htdp]
\begin{center}
\begin{tabular}{lccccccc}
\hline\hline
Galaxy-ID & Stellar disk$^{a}$ & Dust disk$^{b}$ & Extended flux & Fraction of & M$_{D,def.}$$^{d}$ & Fraction of  & M$_{ExtD}$$^{f}$ \\
 & (arcsec) & (arcsec) & (Jy) & the total flux$^{c}$ & $\times$10$^{6}$ (M$_{\odot}$) & dust deficit$^{e}$ & $\times$10$^{6}$ (M$_{\odot}$) \\
\hline \\
\multicolumn{8}{c}{Dusty elliptical galaxies} \\
\hline
HCG04d & 18 & 34 & 0.020 & 17\% & no deficit & 0\%  & 0.16 -- 0.06 \\
HCG15d$^{g}$ & 28 & 19 & -- & -- & 6.44$\pm$2.37 & 76\%  & -- \\
HCG22a$^{g}$ & 43 & 19 & -- & -- & 23.17$\pm$8.55 & 99\%  & -- \\
HCG37e & 11 & 21 & 0.064 & 16\% & 0.09$\pm$0.03 & 3\%  & 0.31 -- 0.12 \\
HCG40a & 28 & -- & -- & -- & 25.52$\pm$9.42 & 70\%  & -- \\
HCG68b$^{g}$ & 55 & 40 & -- & -- & 19.68$\pm$7.26 & 87\%  & -- \\
HCG68d & 25 & 25 & no extend & 0\% & 6.31$\pm$2.33 & 98\% & 0.00 \\
\hline \\
\multicolumn{8}{c}{Normal spiral galaxies} \\
\hline
HCG95c & 12 & 32 & 0.282 & 50\% & no deficit & 0\%  & 4.52 -- 1.62 \\
HCG100c & 24 & 32 & 0.050 & 18\% & 12.68$\pm$6.97 & 9\%  & 0.18 -- 0.07 \\
\hline
\end{tabular}
\end{center}
\label{tab:mass_diffuse_dust}
$^{a}$ Size of the Spitzer/IRAC 3.6$\mu$m stellar disk convolved with the beam of SPIRE 250$\mu$m.\\
$^{b}$ Size of the dusty disk at 250$\mu$m.\\
$^{c}$ Total flux is the sum of the measured fluxes of the stellar disk, plus the diffuse emission.\\
$^{d}$ Dust mass deficiency as described in Sect.~3.1\\
$^{e}$ Fraction of the dust deficit (M$_{D,def}$/[M$_{dust}$+M$_{D,def}$]). \\
$^{f}$ The diffuse dust masses were estimated using Eq.~1 at 250$\mu$m, for $\beta=$1.5 and T ranging between 15-20 K.\\
$^{g}$ Galaxies with dust disk reduced in comparison with the stellar one. \\
$^{h}$ The diffuse dust flux density can be different for HCG92b and HCG92c depending on where we assign the dusty filament.\\
$^{\dagger}$ Galaxy HCG56b is flagged because it host a powerful AGN which dominates its mid-IR colours.
\end{table*}%

\begin{table}[htdp]
\caption{Groups with cold intragroup dust}
\begin{center}
\begin{tiny}
\begin{tabular}{lccccccccc}
\hline \hline
  Group & Dynamical & f$_{60,IRAS}$$^{a}$ & f$_{60,pred.}$$^{b}$ & Fraction of intragroup & f$_{100,IRAS}$$^{a}$ & f$_{100,pred.}$$^{b}$ & Fraction of intragroup & M$_{dust,galaxies}$ &  M$_{D,IGM}$$^{c}$\\
 ID & state &  (Jy) & (Jy) & emission at 60$\mu$m$^{b}$ & (Jy) & (Jy) & emission at 100$\mu$m$^{b}$ & $\times$10$^{6}$ (M$_{\odot}$) & $\times$10$^{6}$ (M$_{\odot}$) \\
 \hline
 HCG16 & DY & 25.09$\pm$0.11 & 26.77 & 6\% & 50.68$\pm$0.06 & 31.03 & 63\% & 64.1 & 204.7 -- 18.1\\
 HCG54 & DY & 0.38$\pm$0.03 & 0.25 & 52\% & 0.71$\pm$0.14 & 0.39 & 82\% & 0.56 & 0.44 -- 0.04\\
 HCG79 & DO & 0.96$\pm$0.09 & 0.41 & 134\% & 2.32$\pm$0.20 & 1.14 & 103\% & 9.33 & 14.9 -- 1.31\\
 HCG92 & DO & 0.58$\pm$0.06 & 0.33 & 76\% & 3.01$\pm$0.30 & 1.26 & 139\% & 31.6 & 39.9 -- 3.48 \\
 \hline
\end{tabular}
\end{tiny}
\end{center}
\label{tab:diffuse}
$^{a}$ Integrated IRAS fluxes from \citet{Allam96} and \citet{Sanders03} for each group, upper limits are not included.\\
$^{b}$ Predicted ``synthetic'' flux for the whole group by adding the corresponding values of all group members estimated using their SED fit.\\
$^{c}$ The intragroup dust masses were estimated using Eq.~1 at 100$\mu$m, for $\beta=$1.5 and T ranging between 15-20 K.\\
\end{table}%

\newpage
\pagebreak
\onecolumn
\appendix
\section{Far-infrared galaxy morphologies}
In this section we present the {\it Spitzer}/IRAC ``true'' colour images of the HCGs in our sample, overlaid with the {\it Herschel}/SPIRE 250$\mu$m cold dust contours in green, as well as with the atomic gas contours (in red; where available). We convolved the {\it Herschel}/PACS and SPIRE bands to the beam of SPIRE 500$\mu$m and compared them to each other to eliminate background sources and mapping artefacts. Therefore, all dust contours observed in these maps are correlated with the groups.  

When examining these maps one can notice extended dust features (i.e. in HCG4, HCG22, and HCG47) and tidal dust bridges (such as the one connecting HCG95C and HCG95A, observed in all bands from the UV to the sub-mm). On the other hand, cases such as HCG22a and HCG68b where their stellar disk extends more than their dust could imply the truncation of the outer dust layers due to tidal stripping. Finally, there are extreme cases, such as HCG92 where all the gas and almost all the dust are located in a large dusty filament outside the main bodies of the galaxies. 

\begin{figure}
\begin{center}
\includegraphics[scale=0.55]{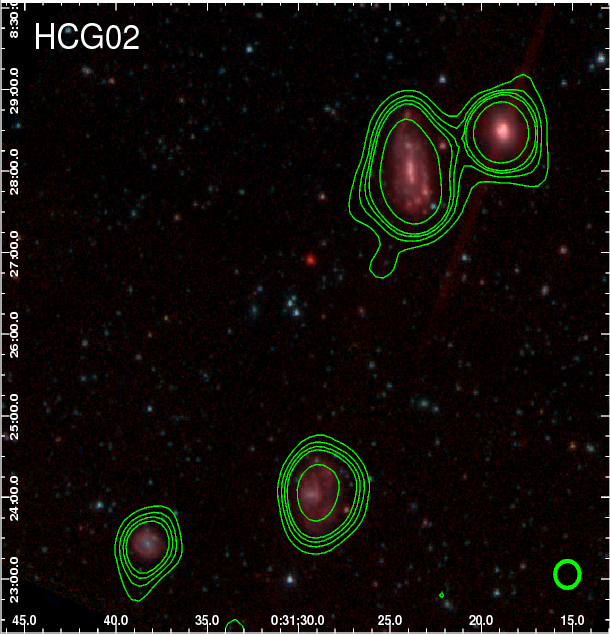}
\caption{True colour images, created using the 3.6, 4.5 and 8.0µm {\it Spitzer}/IRAC mosaics of all the groups in our sample. The green and red (when available) contour maps are the 250$\mu$m {\it Herschel}/SPIRE and the {\it VLA}/HI moment-0 maps of the same groups. The six contours correspond to 3, 5, 7, 9, and 20$\sigma$. In the bottom right corner of each panel we present the size of the SPIRE 250$\mu$m and {\it VLA} band beams, respectively. The RA and Dec scales of the axes are in J2000.0 coordinates.}
\label{fig:Appendix}
\end{center}
\end{figure}

\begin{figure*}
\begin{center}
\includegraphics[scale=0.5]{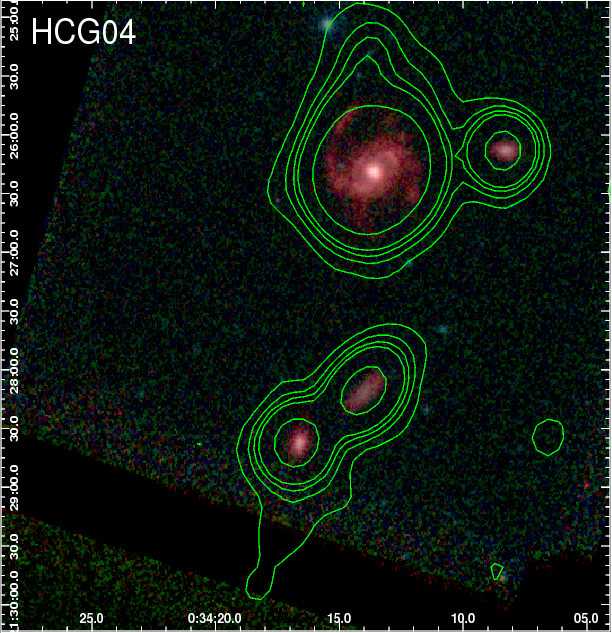}
\end{center}
\end{figure*}

\begin{figure*}
\begin{center}
\includegraphics[scale=0.55]{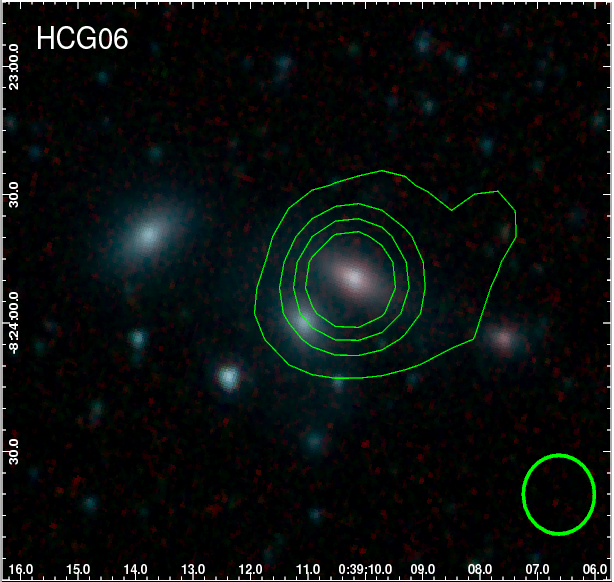}
\end{center}
\end{figure*}

\begin{figure*}
\begin{center}
\includegraphics[scale=0.55]{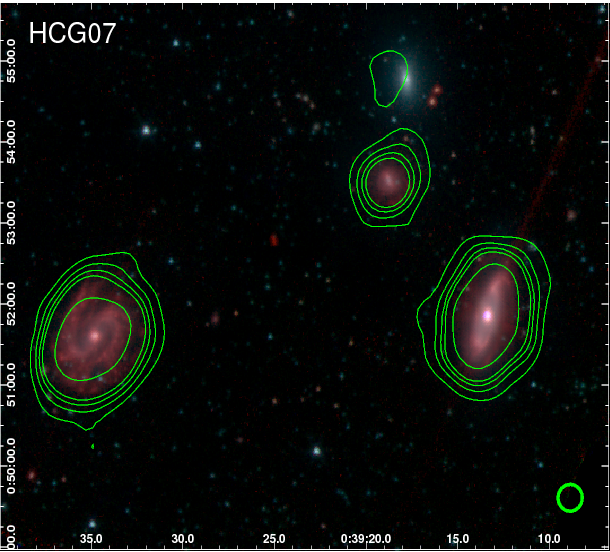}
\end{center}
\end{figure*}

\begin{figure*}
\begin{center}
\includegraphics[scale=0.55]{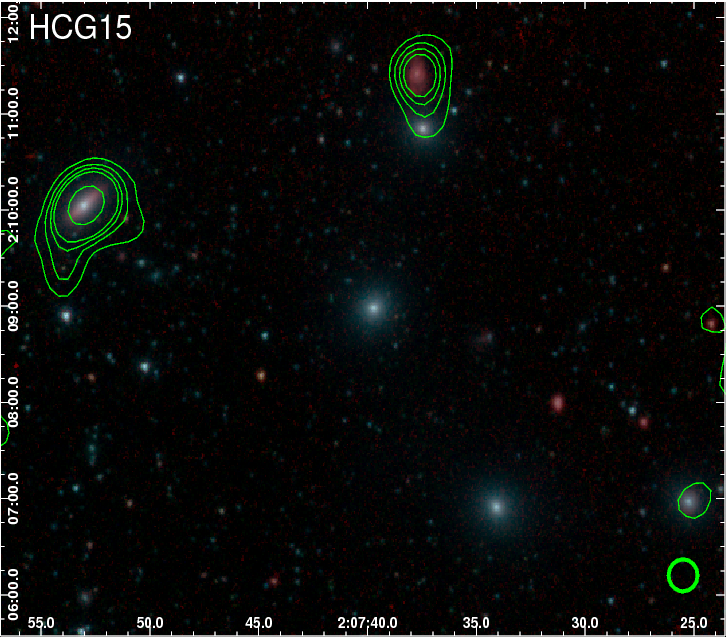}
\end{center}
\end{figure*}

\begin{figure*}
\begin{center}
\includegraphics[scale=0.55]{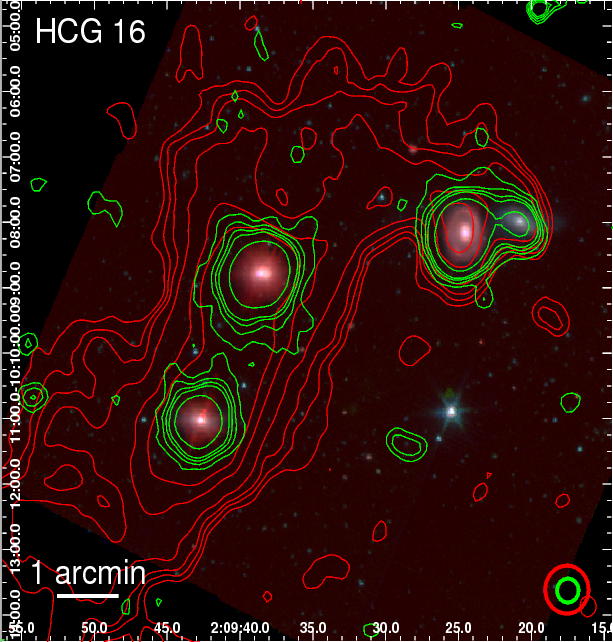}
\end{center}
\end{figure*}

\begin{figure*}
\begin{center}
\includegraphics[scale=0.5]{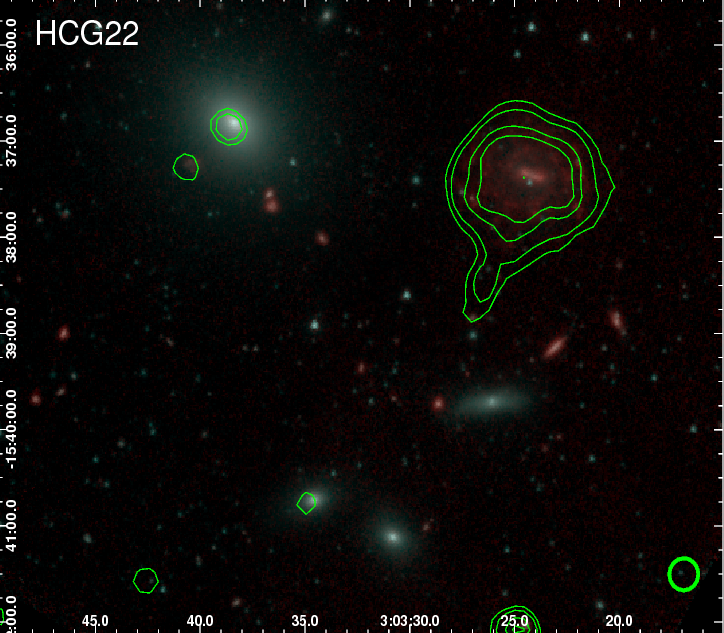}
\end{center}
\end{figure*}

\begin{figure*}
\begin{center}
\includegraphics[scale=0.52]{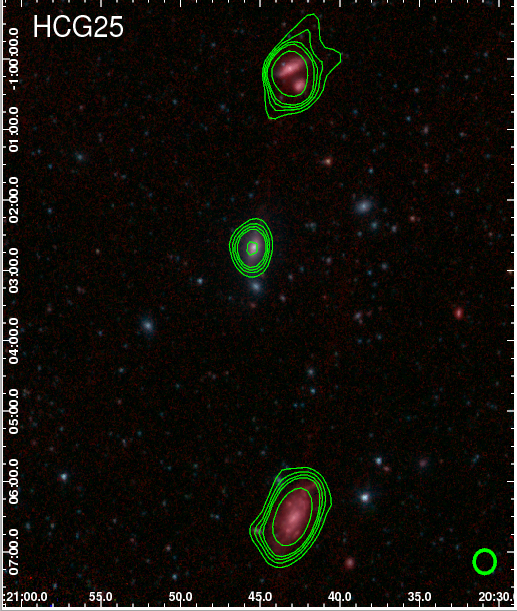}
\end{center}
\end{figure*}

\begin{figure*}
\begin{center}
\includegraphics[scale=0.52]{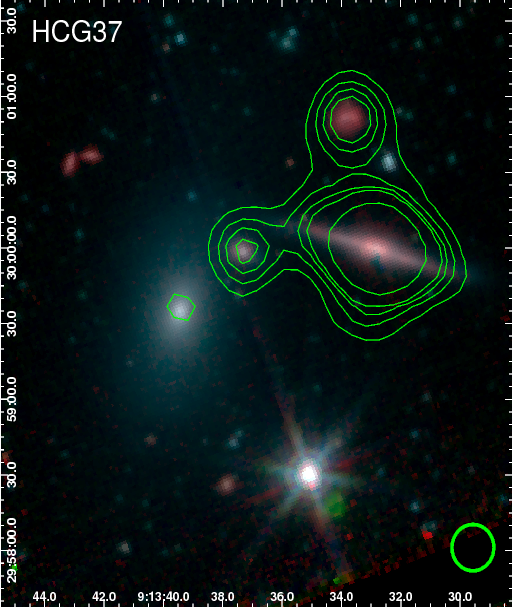}
\end{center}
\end{figure*}

\begin{figure*}
\begin{center}
\includegraphics[scale=0.52]{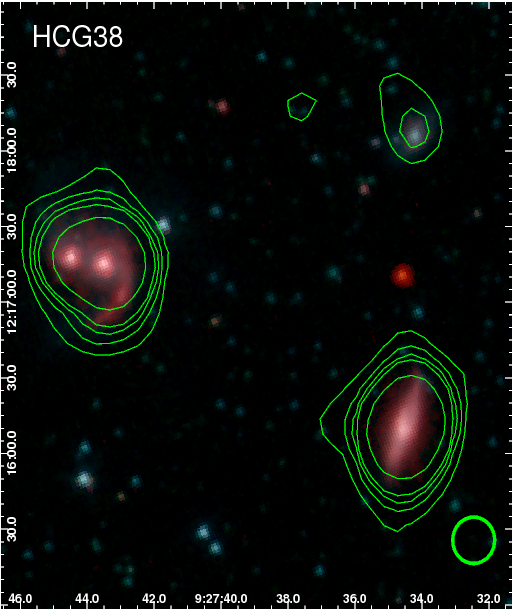}
\end{center}
\end{figure*}

\begin{figure*}
\begin{center}
\includegraphics[scale=0.52]{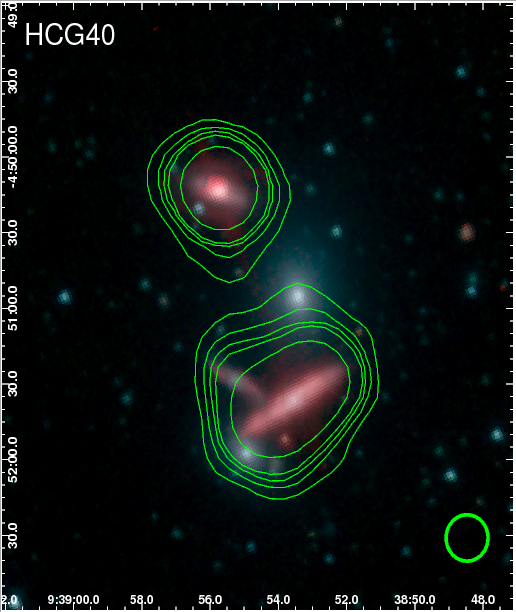}
\end{center}
\end{figure*}

\begin{figure*}
\begin{center}
\includegraphics[scale=0.55]{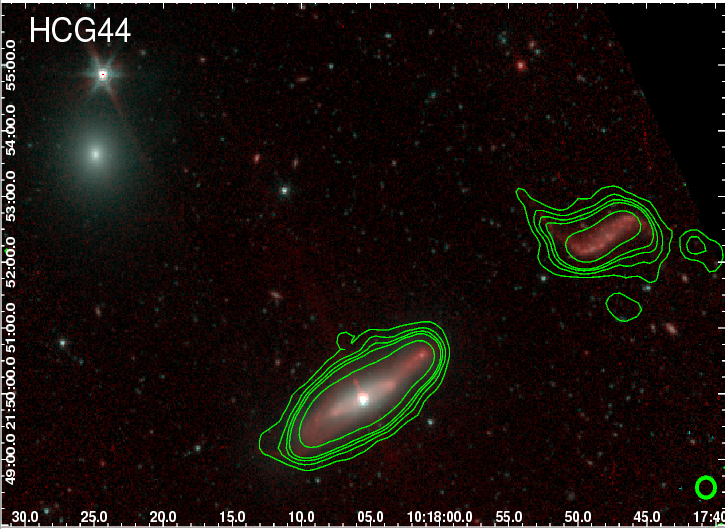}
\end{center}
\end{figure*}

\begin{figure*}
\begin{center}
\includegraphics[scale=0.52]{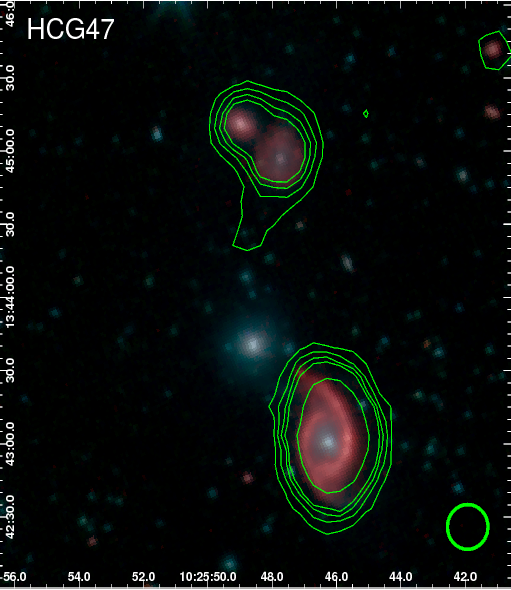}
\end{center}
\end{figure*}

\begin{figure*}
\begin{center}
\includegraphics[scale=0.52]{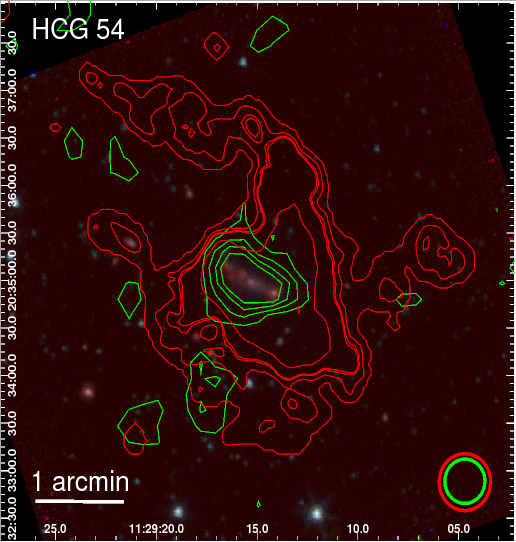}
\end{center}
\end{figure*}

\begin{figure*}
\begin{center}
\includegraphics[scale=0.52]{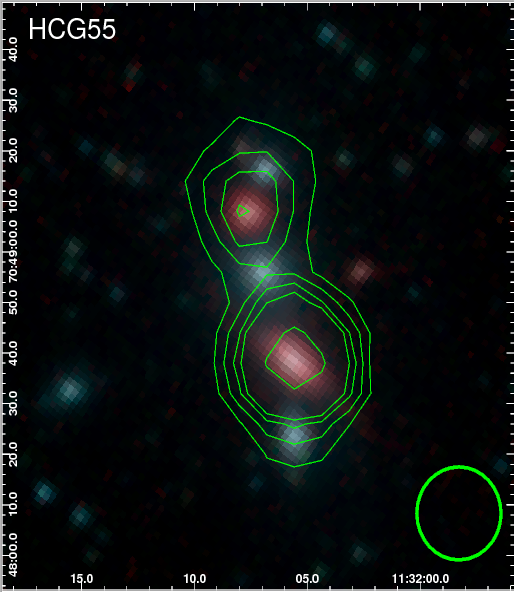}
\end{center}
\end{figure*}

\begin{figure*}
\begin{center}
\includegraphics[scale=0.55]{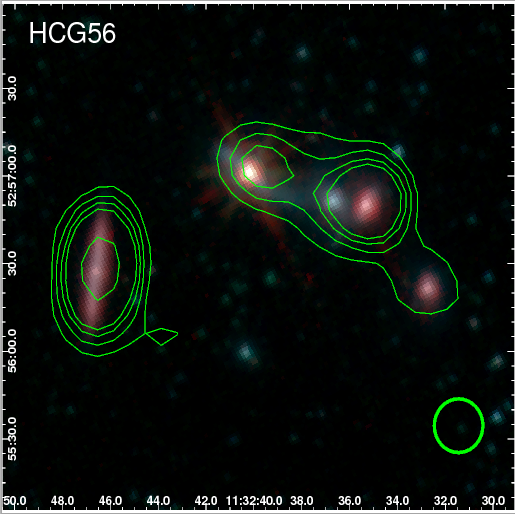}
\end{center}
\end{figure*}

\begin{figure*}
\begin{center}
\includegraphics[scale=0.52]{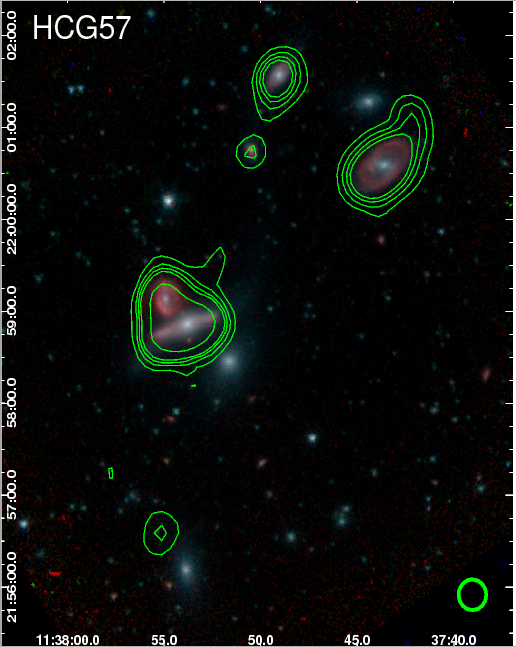}
\end{center}
\end{figure*}

\begin{figure*}
\begin{center}
\includegraphics[scale=0.55]{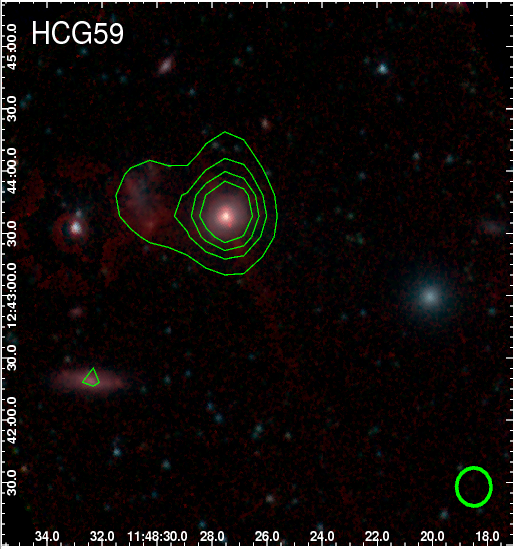}
\end{center}
\end{figure*}

\begin{figure*}
\begin{center}
\includegraphics[scale=0.47]{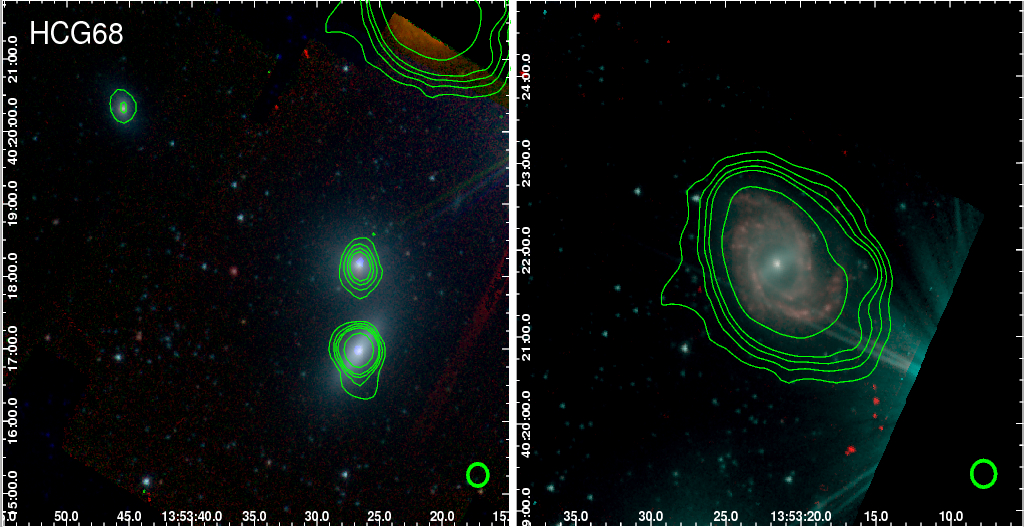}
\end{center}
\end{figure*}

\begin{figure*}
\begin{center}
\includegraphics[scale=0.55]{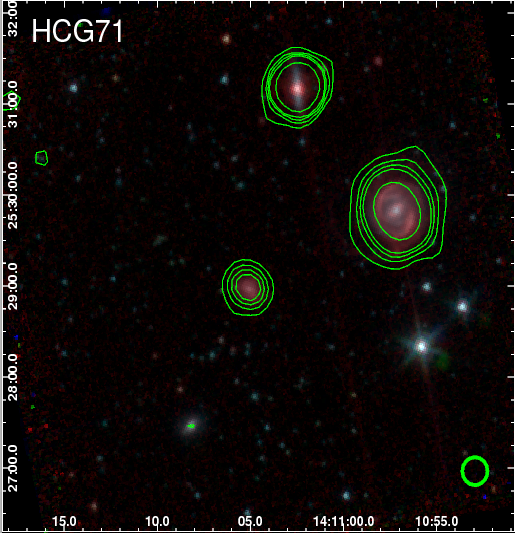}
\end{center}
\end{figure*}

\begin{figure*}
\begin{center}
\includegraphics[scale=0.55]{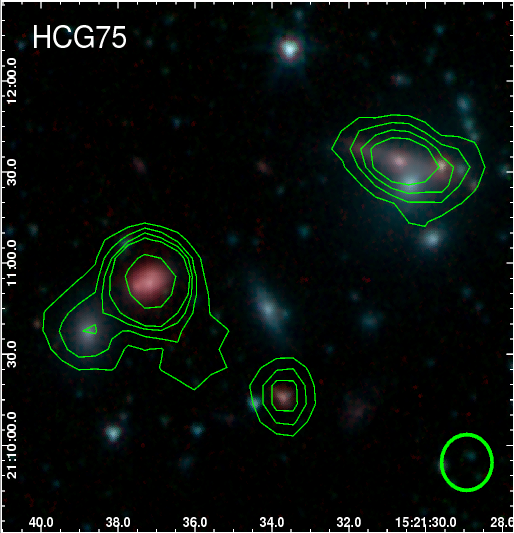}
\end{center}
\end{figure*}

\begin{figure}
\begin{center}
\includegraphics[scale=0.55]{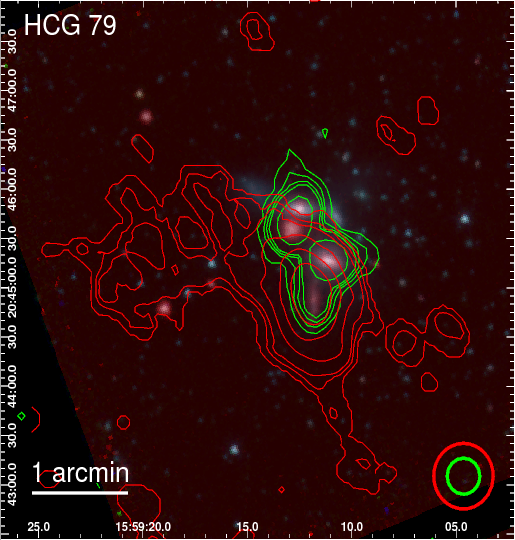}
\end{center}
\end{figure}

\begin{figure*}
\begin{center}
\includegraphics[scale=0.55]{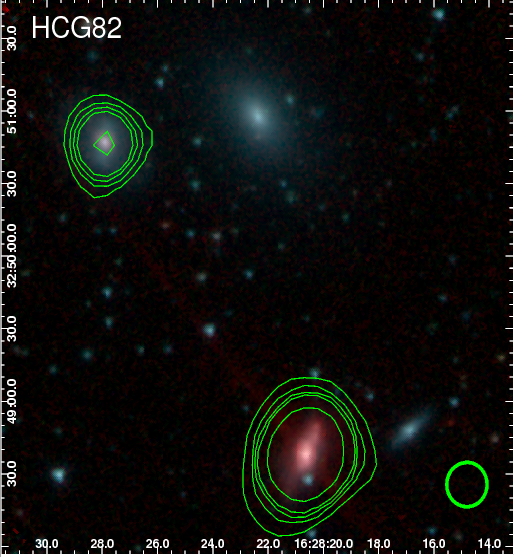}
\end{center}
\end{figure*}

\begin{figure*}
\begin{center}
\includegraphics[scale=0.52]{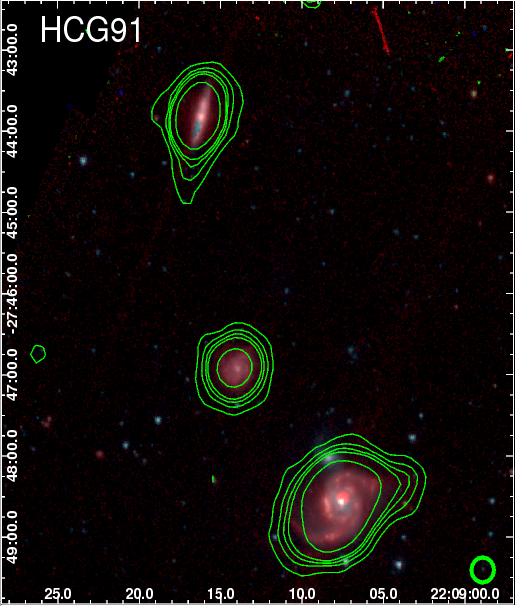}
\end{center}
\end{figure*}

\begin{figure}
\begin{center}
\includegraphics[scale=0.55]{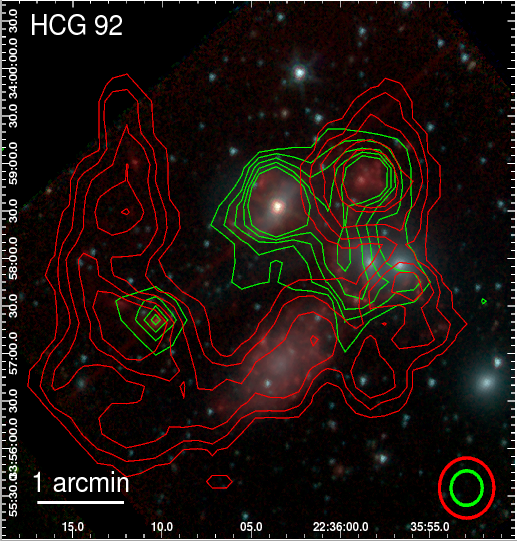}
\end{center}
\end{figure}

\begin{figure*}
\begin{center}
\includegraphics[scale=0.55]{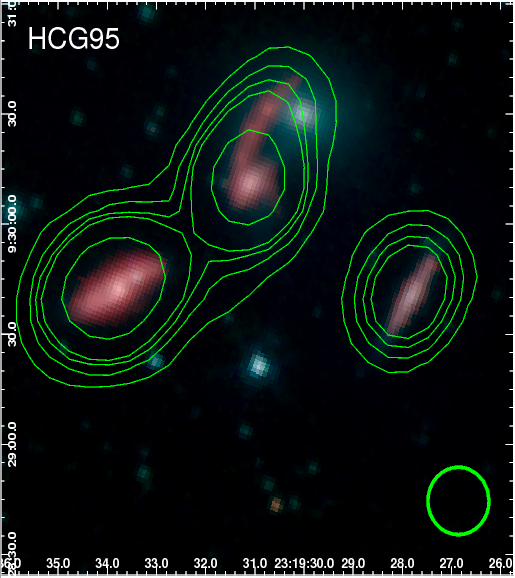}
\end{center}
\end{figure*}

\begin{figure*}
\begin{center}
\includegraphics[scale=0.55]{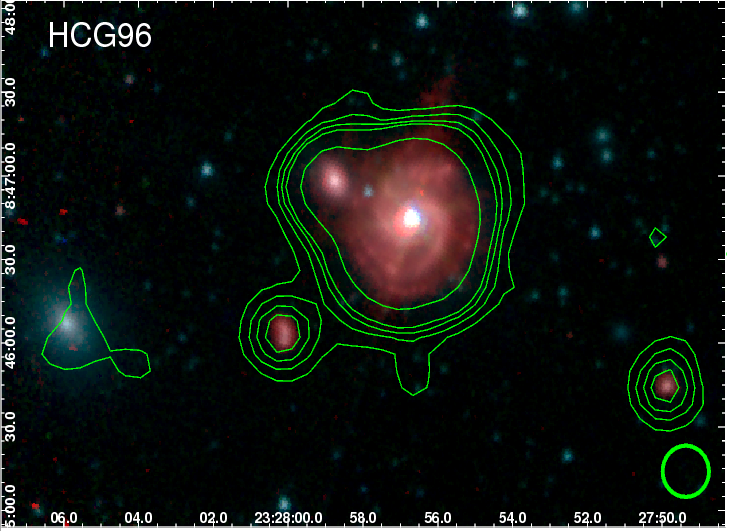}
\end{center}
\end{figure*}

\begin{figure*}
\begin{center}
\includegraphics[scale=0.55]{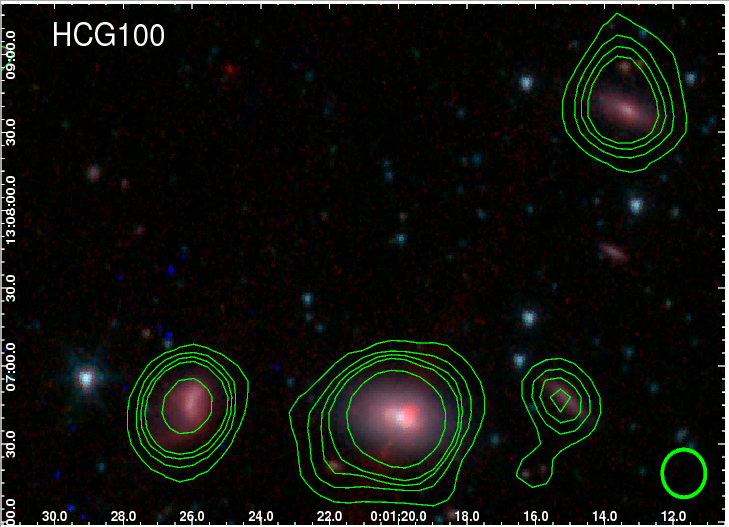}
\end{center}
\end{figure*}

\end{document}